\documentclass[a4paper]{article}                   
\usepackage{graphicx}
\usepackage{natbib}
\usepackage{a4wide}

\newcommand\selen{\textrm{SELEN~}}
\newcommand\selens{\textrm{SELEN}}

\newcommand\sealevel{sea level~}
\newcommand\sealevels{sea level}

\newcommand\jgr{J. Geophys. Res.~}
\newcommand\gjras{Geophys. J. R. Astron. Soc.~}
\newcommand\gji{Geophys. J. Int.~}
\newcommand\grl{Geophys. Res. Lett.~}
\newcommand\eos{Eos. Trans. AGU~}
\begin{document}

\title{Modeling \sealevel changes and geodetic variations by glacial isostasy: the improved \selen code}

\author{Giorgio Spada \\
                 Dipartimento di {Scienze di Base e Fondamenti} (DiSBeF) \\   
                 Universit\`a di Urbino ``Carlo Bo'' \\
                 Urbino, Italy \\
                 \texttt{giorgio.spada@gmail.com} \and 
        Daniele Melini \\
                 Istituto Nazionale di Geofisica e Vulcanologia (INGV) \\
                 Roma, Italy \and
        Gaia Galassi \\
                 Dipartimento di {Scienze di Base e Fondamenti} (DiSBeF) \\   
                 Universit\`a di Urbino ``Carlo Bo'' \\
                 Urbino, Italy \and
        Florence Colleoni \\
                 Centro Euro--Mediterraneo sui Cambiamenti Climatici (CMCC) \\
                 Bologna, Italy
}

\date{\today}

\maketitle


\begin{abstract}
We describe the basic features of \selens, an open source Fortran $90$ program for the numerical solution of the so--called ``Sea Level Equation'' for a spherical, layered, non--rotating Earth with Maxwell viscoelastic rheology. The Sea Level Equation was introduced in the $70$s to model the \sealevel variations in response to the melting of late--Pleistocene ice--sheets, but it can be also employed for predictions of geodetic quantities such as vertical and horizontal surface displacements and gravity variations on a global and a regional scale. \selen (acronym of SEa Level EquatioN solver) is particularly oriented to scientists at their first approach to the glacial isostatic adjustment problem and, according to our experience, it can be successfully used in teaching. The current release (2.9) considerably improves the previous versions of the code in terms of computational 
efficiency, portability and versatility. In this paper we describe the essentials of the theory behind the Sea Level Equation, the purposes of \selens ~and its implementation, and we provide practical guidelines for the use of the program. Various examples showing how \selen can be configured to solve geodynamical problems involving past and present \sealevel changes and current geodetic variations are also presented and discussed. 


\end{abstract}


\section{Introduction}

The physical processes governing \sealevel changes of glacio--isostatic and hydro--isostatic origin have been thoroughly studied in the last decades. They are described in the  so--called ``Sea Level Equation'' (hence after referred to as SLE), the integral equation first introduced by \citet{Farrell_and_Clark_1976} to model \sealevel variations following the melting of late--Pleistocene ice--sheets. The SLE is also currently employed to study the \sealevel changes associated with present terrestrial ice melting in response
to global warming (e.g. \citeauthor{Mitrovica_etal_2001}~\citeyear{Mitrovica_etal_2001}), a critical issue for its potential impact on environment and society \citep{Solomon_etal_2007}. The variation of currently observed geodetic quantities 
(3--D crustal deformations, gravity variations,  altimetric and tide gauges signals) are directly affected by the melting of past and present ice sheets. However, they are also sensitive to the meltwater component of the surface load, whose time evolution is governed by the same equation. The wide spectrum of possible applications of the SLE confirms the central role of glacial isostatic adjustment (GIA) in the framework of both satellite \citep{Peltier_2004} and ground--based geodetic investigations \citep{King_etal_2010}.

\selen is an open source program written in Fortran $90$, primarily designed to simulate the \sealevel and geodetic variations in response to the melting of continental ice sheets. \selen provides a numerical solution to the SLE and gives access to various related quantities of interest for geodynamics and geodesy. The program, which was first introduced in a primitive form by \cite{Spada_and_Stocchi_2007} and now obsolete (hereafter referred to as \selen 1.0), has recently been deeply re--organized and improved in a number of aspects, ranging from the structure of the program to the computational performance and portability. In particular, all components have been reviewed and optimized, and various utilities have been added to facilitate the numerical solution to problems of geodetic relevance, as the time--variations of the harmonic coefficients of the gravity field. Recently, \selen $2.9$ has been successfully employed, for practical applications in the context of GIA, within the {$2$nd Training School on Glacial Isostatic Adjustment Modeling, held in G\"avle (Sweden) in June 2011}. The feedback from the students and from a number of scientists involved in various fields of geodynamics and geodesy, who are routinely working with \selens, have greatly contributed to improve the code previously released by \cite{Spada_and_Stocchi_2007} and has stimulated the release of this upgraded version. 

Though the principal physical ingredients of the SLE are implemented in \selens, some approximations are adopted. First, \selen assumes a linear incompressible rheology and a spherically symmetric undeformed Earth. Therefore, the program takes advantage from the 
viscoelastic Green's function formalism (see~\citeauthor{Peltier_1974}~\citeyear{Peltier_1974} and  
references in \citeauthor{Spada_etal_2011}~\citeyear{Spada_etal_2011}). Consequently, lateral rheological variations are not taken into account. As a second approximation, possible effects of the Earth's rotational fluctuations upon \sealevel variations are not modeled. As discussed by \citet{Milne_and_Mitrovica_1998}, the rotational effects are important in some particular regions of the globe, but not large enough to invalidate the whole solution of the SLE. Third, following \citet{Farrell_and_Clark_1976}, \selen does not account for the horizontal migration of shorelines in response to \sealevel change. This indeed constitutes a crude approximation especially in areas of shallow bathymetry, which will be relaxed in the future releases of \selen so to allow for ``gravitationally self--consistent'' paleogeographical reconstructions \citep{Peltier_2004}. Finally, it is important to remark 
that \selen does not account for tectonic contributions to \sealevels, for local effects such as subsidence driven by sediment loading, nor any possible anthropic contribution to \sealevel change. Furthermore, \selen does not account for ocean
dynamics or possible steric sea{level variations, nor it accounts for 
ice dynamics. The challenging topic of the coupling between the SLE, the ocean circulation and the continental 
cryosphere is left to others.

This manuscript is not self--contained. An exhaustive description of the technical details is avoided here, and emphasis is given on the general features of \selen and the discussion of a set of case studies of general relevance. The theory behind \selen has been illustrated in detail by \citeauthor{Spada_and_Stocchi_06}~(\citeyear{Spada_and_Stocchi_06},~\citeyear{Spada_and_Stocchi_2007}), to which the reader is also referred for references to the most important papers published on this topic. \selen does \textit{not} come with an user guide. However, 
the comments across all Fortran program units and the self--explicatory \selen configuration file should be sufficient. The paper is organized as follows. The physics behind \selen is briefly summarized in Section~\ref{sec:theory}. Section~\ref{sec:how} illustrates how \selen works. Section \ref{sec:case-studies} introduces a few case studies where \selen is employed to model past \sealevel variations (subsection \ref{subsection:rsl}) and present--day variations of geodetic quantities (\ref{subsection:present}). Finally, in Section~\ref{sec:conclusions} we draw our conclusions. 

\selen is free software and anyone is welcome to distribute it under certain conditions within the terms of a Global Public 
License (GPL) (for details, visit \textit{http://www.gnu.org/licenses/}). The \selen package is available upon request to the authors.

\section{Theory}\label{sec:theory}

Here we briefly illustrate the background theory of the SLE, the integral equation that describes the \sealevel 
variations and solid Earth deformations associated with GIA. Essentially, the material that follows is a condensed summary of the 
SLE theory first exposed by \citet{Farrell_and_Clark_1976}, on which \selen is based. In addition, some theory notes 
will also be provided in Section~\ref{sec:case-studies} to discuss specific \selen outputs. For further details, including the 
numerical implementation of the so--called ``pseudo--spectral'' method \citep{Mitrovica_and_Peltier_1991,Mitrovica_etal_1994} 
on which \selen relies upon, the reader is referred to the review of \citet{Spada_and_Stocchi_06} and to references 
therein. 

As it will appear from the synthesis below, the SLE (and its numerical implementation) does not involve absolute quantities and can, consequently, only provide \textit{variations} of geophysical and geodetic quantities relative to a reference state. To clearly illustrate what \selen is 
actually computing, it is convenient to first define ``\sealevels'' as 
\begin{equation}\label{eq:sl-def}
\textrm{SL}(\omega,t) = R_{ss} - R_{se}, 
\end{equation}
where $\omega\equiv (\theta,\lambda)$, $\theta$ is colatitude and $\lambda$ is longitude, $t$ is time, and with 
$R_{ss}$ and $R_{se}$  we denote the radius of the (equipotential) sea surface and of the solid surface 
of the Earth, both relative to the Earth's center of mass, respectively. Quantity $\textrm{SL}(\omega,t)$, which has indeed
an ``absolute'' character, is \textit{not} what \selen is solving for. Rather, the quantity involved in the SLE is 
\sealevel \textit{change}
\begin{equation}\label{sle-original-1}
S(\omega, t) = \textrm{SL}(\omega,t) - \textrm{SL}(\omega,t_{r}), 
\end{equation}
where $t_{r}$ is a reference time that, in the numerical applications of the SLE generally denotes a 
remote epoch prior to the beginning of ice melting. Hence, from (\ref{eq:sl-def}), the definition of \sealevel change 
suitable for GIA studies is  
\begin{equation}\label{sle-original}
S(\omega, t) = N - U, 
\end{equation}
where $N$ represents the sea surface \textit{variation} and $U$ is the vertical displacement of the solid surface of the Earth. Basically, Eq.~(\ref{sle-original}) represents the SLE in its simplest form; 
what follows is aimed to illustrate the relationship between $S$ and the variations of the ice thickness through time,
in order to obtain a form amenable to a numerical approach. 

According to \citet{Farrell_and_Clark_1976}, $N$ can be written as 
\begin{equation}\label{n_sle}
{N}(\omega, t) = G + c,  
\end{equation}
where $c$ is a yet undetermined function of time, and the geoid height variation is 
\begin{equation}\label{g_sle}
G(\omega, t) = \frac{\Phi}{\gamma},  
\end{equation}
in which $\gamma$ is the reference gravity at the surface of the Earth and $\Phi(\omega,t)$ is the total
variation of the gravity potential. Hence, using Eq.~(\ref{g_sle}) into (\ref{sle-original}) gives 
\begin{equation}\label{sle-c}
S(\omega, t) = \frac{\Phi}{\gamma} - U +c, 
\end{equation}
where mass conservation of the system (Ice sheets + Oceans) is ensured taking   
\begin{equation}\label{c-constant}
c = - \frac{m_i(t)}{\rho_{w}A_{o}} - \overline{ \biggl( \frac{\Phi}{\gamma} - U\biggr)}, 
\end{equation}
where the density of water $\rho_{w}$ is assumed to be constant, $m_{i}$ is the mass 
variation of the ice sheets, $A_{o}$ is the (constant) area of the present--day oceans 
and the overbar indicates the average over the surface of the oceans 
\begin{equation}\label{average}
\overline {(\ldots)} = \frac{1}{A_{o}} \int_{o} (\ldots) ~dA,
\end{equation}
where $dA=a^{2}\sin\theta d\theta d\lambda$ is the area element 
and $a$ is Earth average radius. From Eq.~(\ref{sle-c}), the SLE can be therefore written as 
\begin{equation}\label{slegia}
S(\omega,t) = \biggl( \frac{\Phi}{\gamma} - U\biggr)  +S^E   - \overline{ \biggl( \frac{\Phi}{\gamma} - U\biggr)}, 
\end{equation}
where the ``eustatic'' \sealevel variation 
\begin{equation}\label{eustatic}
S^{E}(t) = - \frac{m_i}{\rho_w A_o},  
\end{equation}
shows the remarkable property $S^{E} = \overline{S}$. The SLE has solution $S=S^{E}$ only in the case of a rigid, 
non self--gravitating Earth ($U = \Phi = 0$ in Eq.~\ref{slegia}).

Functions $U(\omega,t)$ and $\Phi(\omega,t)$ will depend on the spatiotemporal variations of the surface load.
This is expressed by  
\begin{equation}\label{load}
{\cal L}(\omega,t) = \rho_{i}I + \rho_{w} S {\cal O},
\end{equation}
where the two terms on the right hand side are associated with the waxing and waning of the ice sheets,
and with the redistribution of meltwater in the ocean basins, respectively. 
In Eq.~(\ref{load}), $\rho_i$ is ice density, ${\cal O}$ is the ``ocean function'' 
(${\cal O}=1$ on the oceans, and ${\cal O}=0$ on land), 
and   
\begin{equation}\label{ice-thick-variation}
I(\omega,t)= T - T_0, 
\end{equation}
is the ice thickness variation, where $T(\omega,t)$ is absolute ice thickness, and $T_{0}(\omega)$ is a reference 
thickness ({e.g.} the thickness at the Last Glacial Maximum, LGM, $21$ kyrs ago). The mass variation in 
Eq.~(\ref{eustatic}) is obtained from (\ref{ice-thick-variation}) by integration over the 
ice--covered regions: 
\begin{equation}\label{mass-variation}
m_{i} (t)= \int_{i} \rho_{i}I ~dA.  
\end{equation}
According to Eq.~(\ref{load}), vertical displacement stems from two terms
\begin{equation}\label{u_sle}
{U}(\omega, t) = \rho_i G_{u}\otimes_i I + \rho_w G_{u}\otimes_o S,  
\end{equation}
where $G_{u}$ is the {Green's} function for vertical displacements, 
$\otimes_i$ and $\otimes_o$ are spatio--temporal convolutions 
over the ice-- and ocean--covered regions, respectively. Similarly
\begin{equation}\label{phi_sle}
\Phi(\omega,t) = \rho_i G_{\phi}\otimes_i I + \rho_w G_{\phi}\otimes_o S,  
\end{equation}
where $G_{\phi}$ is the {Green's} function  
for incremental gravitational potential. Explicit expressions for $G_u$ and $G_{\phi}$ are given in \citet{Spada_and_Stocchi_06}
in terms of the load--deformation coefficients (LDCs) $h(t)$ and $k(t)$, respectively. Introducing the \sealevel Green's function $G_s$ as  
\begin{equation}\label{phi_sle-1}
\frac{G_{s}}{\gamma}(\omega,t) =    \frac{G_{\phi}}{\gamma} - {G_{u}}, 
\end{equation}
substitution of Eqs.~(\ref{phi_sle}) and (\ref{u_sle}) 
into (\ref{slegia}) gives 
\begin{equation}\label{sle}
S(\theta, \lambda, t) = \frac{\rho_i}{\gamma} G_s^{} {\otimes}_i I + \frac{\rho_w}{\gamma} G_s^{} {\otimes}_o S 
                             + S^E 
-\frac{\rho_i}{\gamma} \overline {G_s^{} {\otimes}_i I}
-\frac{\rho_w}{\gamma} \overline {G_s^{} {\otimes}_o S}, 
\end{equation}
which represents the SLE in the ``gravitationally self--consistent'' form that is numerically implemented in \selens. Since the unknown $S(\omega,t)$ also appears in the spatiotemporal convolutions at the right--hand side, the SLE is an integral equation, which cannot be solved explicitly unless some drastic simplifying assumptions are made \citep{Spada_and_Stocchi_06} . The SLE is a linear equation as long as shorelines are not allowed to migrate horizontally, i.e. if ${\cal O}$ (and consequently $A_o$) are not time--dependent. Sea level variations are sensitive to mantle rheology through $G_{s}$, since this is determined by the viscoelastic LDCs $h$ and $k$ \citep{Spada-2003, Spada_and_Stocchi_06}. Solutions of Eq.~(\ref{sle}) in special cases, discussed in detail by \citet{Spada_and_Stocchi_06}, are also available via \selen. 

\section{How \selen works}\label{sec:how}

Running \selen requires a standard UNIX (including Linux and Mac OS X) environment and a Fortran 90 compiler. On Windows systems, \selen can run within the Cygwin environment, freely downloadable from 
\textit{http://www.cygwin.com}. A list of supported operating systems and Fortran compilers is given in the (simple text) configuration file \texttt{config.dat}. 
While the previous version of \selen (1.0) was limited to the IBM XL commercial compiler, \selen 2.9 components can be compiled both using the freely available g95 (\textit{http://www.g95.org}) and gfortran (\textit{http://gcc.gnu.org/gfortran})
compilers, or the commercial Intel Fortran compiler. Additional configurations for the operating system and compilers in 
\selen 2.9 can be implemented by modifying the setup program \texttt{config.f90}. To run \selens, the GMT (Generic Mapping Tools) public domain software of \citet{Wessel_and_Smith_1998} must be installed on the system. In \selen 2.9, the most computationally intensive portions of code have been parallelized with~\cite{Openmp_2005} directives. The corresponding program units can therefore take advantage of multi--threading on modern CPUs, resulting in a substantial performance improvement. 

\selen consists of several (independent) Fortran 90 program units, performing specific computation steps. A list, with a short 
description of their purpose, is provided in \marginpar{T\ref{table:some-files}} Table~\ref{table:some-files}. The user can specify run parameters in the configuration file \texttt{config.dat}; this file is parsed by the setup program \texttt{config.f90}, which checks parameters for consistency and creates a shell script (\texttt{selen.sh}) containing the compilation and execution commands for the \selen components
needed for the specific run. 
The configuration and execution process is transparent to the user and entirely handled by the \texttt{makeselen.sh} shell script. To launch a \selen run, the user only needs to edit the \texttt{config.dat} file and execute the \texttt{makeselen.sh} script. {This 
constitutes a significant improvement with respect to previous version \selen 1.0, which lacked a flexible user interface.  
After each run, \selen outputs are stored in subfolders of the output folder named \texttt{depot-}\textit{name}, created in the working directory, where \textit{name} is a 4--character label defined by the user in the \texttt{config.dat} file.  

In this way, outputs of different \selen runs can be stored in the working directory.  The structure of the output folder is described in \marginpar{T\ref{table:folders}} Table~\ref{table:folders}; a copy of the \texttt{config.dat} file is also stored in the output folder for reference.

A typical \selen run can be subdivided in three main steps: \textit{i)} pre--computing of all the needed functions, \textit{ii)} numerical 
solution of the SLE and \textit{iii)} computation of geophysical and geodetic quantities requested by the user. 

During the first phase, \selen performs the spherical harmonic (SH) expansions of the ocean function ${\cal O}$ and of the 
ice thickness variation $I$, and obtains the LDCs $h$, $l$ and $k$ for the selected Earth model (an Earth model is defined 
by its elastic and rheological layering according to the conventions adopted for the TABOO program, see 
\citeauthor{Spada-2003}~\citeyear{Spada-2003}). For the numerical evaluation of surface integrals of Eq.~(\ref{slegia}), \selen defines an icosahedron--shaped pixelization of the sphere, displayed in Fig. \ref{fig:pixelization}. \marginpar{F\ref{fig:pixelization}} The grid density is controlled by the resolution parameter $R$, defined in the \texttt{config.dat} file; the number of grid pixels  is given by $N_p = 40R(R+1) + 12$. As discussed by \cite{Tegmark_1996}, an accurate numerical integration is only ensured for $\ell_{max} \le \sqrt{3N_p}$, where $\ell_{max}$ is the maximum harmonic degree of the SH expansion.  A discrete realization of the ocean function is practically obtained by selecting the grid pixels falling in ``wet'' and ``dry'' areas with the \texttt{gmtselect} utility from the GMT package (see 
Fig.~\ref{fig:pixelization}). To optimize computations involving SH functions, all the needed associated Legendre functions and trigonometric functions at grid pixels are pre--computed once and stored on disk for later re--use.  This SH database can be re--used also across different  runs, as long as parameters $R$ and $\ell_{max}$ are unchanged (values of $R$ and $\ell_{max}$ are defined in the \texttt{config.dat} file). Similarly, the SH expansion of the ice load, a relatively time--consuming task, can be re--used by setting the appropriate option in the configuration file \texttt{config.dat}.  

In the second phase, \selen solves numerically the SLE. As discussed in Section \ref{sec:theory}, the SLE is an integral (implicit) equation, in which the unknown $S$ is convolved in space and time with the sea level Green's functions. An iterative solution scheme is therefore needed, similar to the one employed in the solution of one--dimensional inhomogeneous integral Fredholm equation of the second kind. As a zero--order solution, \selen approximates the unknown function $S$ with the eustatic sea level variation: $S^{(0)} = S^E$. By substituting $S^{(0)}$ in the right--hand side of Eq.~(\ref{sle}), a new estimate $S^{(1)}$ is obtained. This process can be iterated any number of times, by using Eq.~(\ref{sle}) to obtain the $k$-th order solution $S^{(k)}$ from the $(k-1)$--th order solution $S^{(k-1)}$, until a pre--determined convergence threshold 
is reached (i.e. until the ratio $|S^{(k)}-S^{(k-1)}|/|S^{(k)}|$ becomes sufficiently small at all the grid pixels and time steps). According to the numerical tests carried out by \cite{Spada_and_Stocchi_2007} and to our experience, three iterations normally suffice for convergence.

In the third and final phase, \selen uses the solution of the SLE in the spectral domain to obtain predictions of geophysical and geodetic quantities by SH synthesis. According to the settings in the user configuration file \texttt{config.dat}, \selen can compute present--day rates of deformation (in vertical and horizontal directions), geoid change and \sealevel change on global or regional scales, relative sealevel predictions at specific sites, \sealevel change rates at tide gauge stations,  present--day velocities at user--specified sites, and present--day rates of variation of the Stokes coefficients of the Earth's gravity field. 

The execution time of \selen scales with spatial and harmonic resolutions as $t_{exe} \sim N_p N_h$, where $N_p$ is the number of grid pixels and $N_h = (\ell_{max}+1) (\ell_{max}+2) / 2$ is the number of SH functions with harmonic degree $\ell\le \ell_{max}$; as a rough approximation, $t_{exe} \sim R^2 \ell_{max}^2$, where $R$ is the resolution parameter. The memory footprint of \selen follows a similar scaling law. For high resolution runs (large values of $\ell_{max}$ and consequently of $R$), disk I/O can become a considerable fraction of $t_{exe}$, and these relations may no longer be valid. On multi--core systems, \selen can use multi--threading to reduce computation time (multi--threading was not implemented in previous version \selen 1.0). To enable multi--threading, the corresponding option in \texttt{config.dat} must be set, specifying the
number of threads that \selen will create. The number of threads should be equal to the number of cores in the system; it can be smaller if the user wishes to leave some resources free for other tasks.

\section{A few case studies}\label{sec:case-studies}

To illustrate the main features of \selens, we have configured the program for a GIA simulation 
based on model ICE--5G \citep{Peltier_2004}. In the previous version (\selen 1.0) 
this ice model, which is now widely 
employed in the GIA literature, was not accessible. The configuration file \texttt{config.dat} for this run
(referred to as \texttt{TEST} run) and all the \selen output files (text files and plots) are available 
in the folder \texttt{depot-TEST} that comes with the \selen distribution package.  The numerical results 
presented below are organized into two subsections. In the first (Section \ref{subsection:rsl}), the focus is 
on the Relative Sea Level (RSL) variations  driven 
by GIA since the LGM, $21$~ kyrs ago. The second (Section~\ref{subsection:present}) 
is about the delayed viscoelastic effects that GIA is still producing today, and the focus is on the time variations
of various geodetic quantities. Keeping separate the two time scales helps to simplify the presentation of the results. 
However, it should be kept in mind that the present--day geodetic variations depend on the whole history of deglaciation, 
and are thus extremely sensitive to mantle viscosity in spite of the long time elapsed since the end 
of melting of major continental ice sheets ({e.g.} \citeauthor{Peltier_2004}~\citeyear{Peltier_2004}). 

The viscosity profile employed in the cases studies illustrated below is a three--layer volume--average 
of the original, multi--layered ``VM2'' profile introduced by \citet{Peltier_2004} and associated with the ICE--5G model. 
Values of viscosity 
and other parameters that define the incompressible Earth model (hereafter referred as to VM2a) 
are summarized in \marginpar{T\ref{table:vm2}} Table~\ref{table:vm2}. The isostatic relaxation spectrum 
determined by \selen for VM2a is shown in \marginpar{F\ref{fig:spectrum}} Fig.~\ref{fig:spectrum}, while \marginpar{F\ref{fig:ldcs}} Fig.~\ref{fig:ldcs} shows the elastic and fluid LDCs. \selen incorporates 
TABOO \citep{Spada-2003,tabooeos} as a subroutine for computing these spectra, based on the Viscoelastic Normal Modes theory \citep{Peltier_1974}. After execution, all the information about the LDCs (tables with ASCII data 
and plots, if requested) are made available to the user in folder \texttt{depot-TEST/Love-Numbers-by-TABOO}, including the 
viscous parts of the LDCs (not shown here). All the ensuing \selen results are obtained using LDCs in the 
range of degrees $1 \le \ell_{max} \le 128$ and are expressed in the reference frame with origin the Earth 
center of mass (hence, differently than in \selen 1.0, here we properly account for the degree $\ell=1$ LDCs). 
LDCs of harmonic degree $\ell=0$ vanish identically because of incompressibility, but they would not play any role 
even for an compressible Earth since the SLE includes explicitly the constraint of mass conservation (see Eq.~\ref{c-constant}). 
The SLE is solved iteratively on the  grid shown in Fig.~\ref{fig:pixelization} 
and three iterations are performed to ensure convergence of the iteration scheme \citep{Spada_and_Stocchi_2007}. 

The spatial distribution of the ice thickness according to model ICE--5G at the LGM and at present time are shown in the \selen plots of \marginpar{F\ref{fig:ice5g}} Fig.~\ref{fig:ice5g}. The ICE--5G thickness data are obtained from the home page of Prof. W. R. Peltier (the version with $1^\circ$ resolution is used). In folder \texttt{ICE-MODELS} of the main \selen directory, other ice models are available, including the previous ICE--X models developed by Prof. W. R. Peltier and co--workers (\selen also incorporates individual components of the global ICE--5G model, which could be useful for regional studies). Using specific {Fortran} formats described in program \texttt{config.f90}, the user can indeed introduce other \textit{ad hoc} ice models according to specific purposes. In the specific case of ICE--5G, the spatiotemporal discretization has been slightly modified for the sake of computational convenience (details on the discretization scheme adopted in \selen are available in \citeauthor{Spada_and_Stocchi_2007}~\citeyear{Spada_and_Stocchi_2007}). The original ICE--5G time grid has been converted into a uniformly spaced $1$--kyr grid, and the elementary ``rectangular'' $1^{\circ} \times 1^{\circ}$ ice elements that define ICE--5G are converted, at run time, into equal--area and equal--thickness ``discs'' that allow for a straightforward SH decomposition because of their symmetry (see {e.g.} \citeauthor{Spada_and_Stocchi_06}~\citeyear{Spada_and_Stocchi_06}). Since in \selen $2.9$ fixed shorelines are assumed, the melting of the ice distributed across the continental shelf in Fig.~\ref{fig:ice5g} (top) is not accompanied by a transgression of ocean water, which may imply a local error in \sealevel change predictions (this restriction will be relaxed in future releases of \selens). The equivalent \sealevel (ESL), shown in \marginpar{F\ref{fig:eustatic}} Fig.~\ref{fig:eustatic}, provides the time--history of the ice volume of ICE--5G since the LGM. After execution, 
all the ice model data are stored in \texttt{depot-TEST/ICE5G}. These include the spatial distribution of the ice masses at all
times, the ESL function, the SH coefficients of the ice thickness and the SH reconstruction of the ice distribution (options for these computations are available in file \texttt{config.dat}). 

\subsection{Past Relative Sea Level variations}\label{subsection:rsl}

One of the purposes of \selen is the modeling of \sealevel variations that occurred 
during the time period elapsed since the LGM, in consequence of the melting of late--Pleistocene 
ice sheets. There is a considerable amount of literature about this problem, and it is impossible 
to provide an exhaustive summary in this paper. A flavor of the enormous importance that
past \sealevel variations have on our understanding of present--day variations 
is given by {e.g.} \citet{Lambeck_and_Chappell_2001} and \citet{Peltier_2004} and references therein. 

In this section, we provide an illustration of the \selen outputs regarding ``Relative Sea Level''
(RSL).  This quantity is not directly obtained from the SLE (\ref{sle}), which provides $S(\omega,t)$. 
According to the geological practice, RSL is defined as the difference
between \textrm{SL} at a given epoch before present ($t=t_{BP}$) and the present--time ($t=t_p$) value: 
\begin{equation}
\textrm{RSL}(\omega,t_{BP})=\textrm{SL}(\omega,t_{BP})- \textrm{SL}(\omega,t_{p}),
\end{equation}
which, using the definition of \sealevel given by Eq.~(\ref{eq:sl-def}), can be also written as  
\begin{equation}
\textrm{RSL}(\omega,t_{BP})= {S}(\omega,t_{BP})- {S}(\omega,t_{p}),
\end{equation}
showing that $\textrm{RSL}$ can indeed be obtained from the solution of the SLE computed at two 
different times (we note that $\textrm{RSL}$ does not depend on the choice of the remote time 
$t_{r}$ in Eq.~(\ref{sle-original-1})). All the following RSL computations have been obtained 
performing three iterations within the solution scheme of the SLE, which generally 
provides a sufficiently accurate solution (see the test computations in \citeauthor{Spada_and_Stocchi_2007}
\citeyear{Spada_and_Stocchi_2007}). 

\selen can be easily configured for producing RSL predictions at sites of interest to the user, 
for which observations of past sea levels can be obtained from the literature. 
\marginpar{F\ref{fig:rsl-sites}} Fig.~\ref{fig:rsl-sites}, produced 
by \selens, shows the locations of $392$ sites for which radiocarbon--controlled RSL data are available, 
according  to the compilation of \citet{Tushingham_and_Peltier_1992,Tushingham_and_Peltier_1993}. Information 
on the sites coordinates and RSL observations, including their uncertainties, are contained in file 
\texttt{sealevel.dat}, obtained from the National Oceanic and Atmospheric Administration (NOOA) page 
\footnote{See \textit{http://www1.ncdc.noaa.gov/pub/data/paleo/paleocean/ 
relative\_sea\_level/.}}. The RSL observations in this file cover the last $15,000$ years and are 
obtained from various sources, mainly based on geomorphologic and archaeological methods. 
In the scatterplot of \marginpar{F\ref{fig:rsl-scatterplot}} Fig.~\ref{fig:rsl-scatterplot}, all RSL
observations from the input file \texttt{sealevel.dat} (top frame) are qualitatively compared with predictions
obtained from \selen using the settings in \texttt{config.dat} (bottom). The scatterplot gives a clear 
view of the temporal distribution of the RSL observations from this data collection, which are mainly relative 
to the last $\sim 8,000$ years. The similarity of the two scatterplots in this figure clearly indicates that 
the ICE--5G model, 
in its \selen implementation, broadly reproduces the RSL observations globally available. Of course, local
misfits are possible, as we will discuss below for specific sites. A more rigorous global misfit analysis 
is also possible using \selens, oriented to users interested to tackle an optimization problem in which the 
best fitting model parameters (i.e. the viscosity of mantle layers) are to be inverted from a specific set 
of RSL observations (the results of the misfit analysis for this run, not reproduced here, are available 
in the folder \texttt{depot-TEST/rsl/rsl-misfit}). 

Fig.~\ref{fig:rsl-hbay} \marginpar{F\ref{fig:rsl-hbay}} shows a RSL analysis, in which \selen is 
programmed to produce RSL predictions at individual sites. Here we only show eight curves out of the $392$ 
available in folder \texttt{depot-TEST/rsl/ rsl-curves} after the execution of \selens, all relative to 
the case of the Hudson Bay. The process of GIA across the Hudson 
Bay has been the subject of various investigations in the past, because of the sensitivity shown by the uplift 
data on the rheological layering of the mantle \citep{Mitrovica_and_Peltier_1992,Cianetti_etal_2002}. RSL observations 
from this region, shown by error bars in Fig.~\ref{fig:rsl-hbay}, clearly indicate a monotonous \sealevel
fall that is typical of sites belonging to previously ice--covered areas during the LGM. 
Model predictions reproduce the RSL observations very satisfactorily. This should not come as a surprise,
since the two basic components of model ICE--5G (namely, the chronology of the late--Pleistocene 
ice sheets and the viscosity profile of the mantle) have been expressly designed to best fit a global dataset 
of RSL observations, which includes these Hudson bay sites \citep{Peltier_2004}. The further example of Fig.~\ref{fig:rsl-var} 
\marginpar{F\ref{fig:rsl-var}} shows results for eight RSL sites in the far field of the previously ice--covered regions, 
where the history of \sealevel rise has been less influenced by the direct effect of ice melting. The shapes
of the RSL curves in these regions clearly depart from those of the Hudson bay in Fig.~\ref{fig:rsl-hbay}. In some
cases, they neatly indicate a monotonous \sealevel rise, but some remarkable exceptions are found,
in which they show \sealevel high--stands or more complex features. Their trend is reasonably reproduced 
by our \selen computations. 

In \marginpar{F\ref{fig:rsl-sleapp}} Fig.~\ref{fig:rsl-sleapp} we show the results of further runs, 
in which the program has been configured to solve the SLE using various approximations. The ice model 
and the rheology are the same as in previous computations. 
This kind of analysis, which is performed here only for the sites of Richmond (Hudson bay) and of Merseyside (England), 
is not meant to test the agreement of RSL predictions with observations. Rather, it 
can be useful to illustrate the role played by some of the basic physical ingredients of the SLE. 
The solid curves still show the ``gravitationally self--consistent'' (GSC) solution, in which the SLE is solved in the full form given by Eq.~(\ref{sle}). In results shown by dotted RSL curves, only the elastic (EL) components of the LDCs are employed in the computations, hence neglecting the viscous components of the Maxwell rheology. The eustatic curve (EU, dashed), shows the RSL trend obtained assuming a rigid Earth and neglecting any gravitational interaction between the solid Earth, the ice masses, and the oceans. The eustatic RSL curve, which is simply expressed by 
Eq.~\ref{eustatic}, is the same for all RSL sites. Finally, the dash--dotted curve (W) shows the solution of the SLE in the so--called ``Woodward approximation'' (see \citeauthor{Spada_and_Stocchi_2007} \citeyear{Spada_and_Stocchi_2007}), in which the Earth is assumed to be perfectly rigid (LDCs are $h=l=k=0$), and only the gravitational attraction between the ice masses and the oceans is 
taken into account. From the results of Fig.~\ref{fig:rsl-sleapp}, it is apparent that the RSL observations across the Hudson bay, where ice melting has produced a huge uplift, can only be explained invoking the ``gravitationally self--consistent'' solution. However, in the case of the far--field site Merseyside, the GSC solution can be assimilated to any of the approximate solutions (including those pertaining to a rigid Earth, EU and W), at least for the last few kiloyears. 

The site--by--site analyses shown in Figs~\ref{fig:rsl-var} and \ref{fig:rsl-sleapp} above are useful 
for a detailed reconstruction of the local RSL curves and direct comparison with observations. However, 
a global or regional visualization of the RSL variations can be more useful for a qualitative interpretation. 
For the purpose of visualization of the relative \sealevel variations associated with GIA, 
\selen can be configured in two different ways. Both features were not directly accessible to the user in 
\selen 1.0. The first, which is oriented to global 
analyses, allows for the visualization of the so--called ``Clark zones'' \citep{Clark_etal_1978}, i.e. 
the regions of the globe in which the RSL curves shows similar patterns after the LGM.  
Numerical results and plots for this analysis, which are not reported here for space limitations, will 
be accessible for the \texttt{TEST} run in folder \texttt{depot-TEST/rsl/rsl-zones/} after 
execution of \selens. An example of the second possible RSL analysis, which is more oriented to regional 
investigations, is shown in  \marginpar{F\ref{fig:med-rslc}} Fig.~\ref{fig:med-rslc}, where RSL contour 
plots across the Mediterranean region for times $2$, $6$ and $10$ kyrs BP are drawn. This area is the subject of considerable
interest in view of the significant amount of \sealevel indicators available 
\citep{Lambeck_etal_2004,Lambeck_and_Purcell_2005}. Fig.~\ref{fig:med-rslc} provides a clear regional
characterization of the process of \sealevel rise in the area. Although the RSL pattern shows a 
considerable complexity, it is apparent that the largest RSL excursions are predicted 
in the bulk of the basin, in consequence of the broad subsidence caused by the effects of meltwater load 
\citep{Stocchi_and_Spada_2007}. Data and maps for this kind of analysis are found in folder 
\texttt{depot-TEST/rsl-contours/}.   
 
\subsection{Effects of GIA on present--day geodetic variations}\label{subsection:present}

Three--dimensional movements of the solid Earth and gravity variations are still affected significantly 
by the isostatic disequilibrium in response to the melting of late--Pleistocene ice sheets (see {e.g.} \citeauthor{King_etal_2010}~\citeyear{King_etal_2010}). 
Therefore, it is sometimes necessary to evaluate quantitatively the amplitude of the 
GIA effects at present time, possibly in order to decontaminate geodetic quantities observed locally or globally. 
Since for loads of the size of major late--Pleistocene ice sheets the Maxwell 
relaxation time of the mantle is of the order of a few kilo--years (see {e.g.} 
\citeauthor{Schubert_etal_2004}~\citeyear{Schubert_etal_2004}), the  
rates of \sealevel change or other quantities associated with GIA can be effectively considered 
as constant through a decade to century time scale.   
Hereinafter, we will consider a few examples in which the SLE is solved for trends of GIA--related
quantities. Since solving the SLE implies a temporal discretization of all variables involved (see 
\citeauthor{Spada_and_Stocchi_2007}~\citeyear{Spada_and_Stocchi_2007}  
for details), in \selen geodetic trends at a specific place $\omega$ are evaluated numerically as 
\begin{equation}\label{eq:derivative}
\dot Q \equiv \frac{dQ}{dt}(\omega,t_p)  \approx \frac{Q(\omega,t_{p})-Q(\omega,t_{p}-\Delta)}{\Delta}, 
\end{equation}
where $\Delta=1$ kyrs is the natural time step in \selens, 
and $Q$ here represents any of the fundamental geodetic quantities $S$ (relative sea level variation), $U$ 
(vertical disaplacement) and $N$ (sea surface variation) that appear in Eq.~\ref{sle-original}. Of course, 
more accurate methods can be implemented by the user for the numerical evaluation
of trends. Thought here we focus in the effects of melting of past ice sheets on present--day 
geodetic variations, with minor modifications \selen can be also employed for the study of the
fingerprints \citep{Mitrovica_etal_2001,Tamisiea_etal_2001} of the current terrestrial ice melting. Examples have been recently given
by \citet{Sorensen_etal_2011} and \citet{Spada_etal_2012} for the case study of the melting of the Greenland ice sheet. 

Present--day trends $\dot S$, $\dot U$ and $\dot N$, computed by means of Eq.~(\ref{eq:derivative}), 
are shown in \marginpar{F\ref{fig:sdot}} Figs~\ref{fig:sdot}, \marginpar{F\ref{fig:udot}} \ref{fig:udot} and 
\marginpar{F\ref{fig:ndot}} \ref{fig:ndot}, respectively, for our \texttt{TEST} run. 
Of course, these global maps are not independent of one another, since from the SLE 
(see Eq.~\ref{sle-original}), we have e. g. $\dot N=\dot S + \dot U$. To emphasize the 
regional variations of these GIA signals, the color tables span the range of $\pm 1$ mm/yr. 
The very complex pattern of \sealevel change $\dot S$ in Fig.~\ref{fig:sdot} is the consequence 
of the delayed visco--elastic deformations of the solid surface of the Earth, of gravitational 
interactions between the solid Earth, the ice masses and the oceans, and of the complex 
shape of continents. For a rigid, non gravitating Earth we would observe $\dot S=0$ everywhere 
on this map. The $\dot S$ pattern in Fig.~\ref{fig:sdot} has been discussed in a number of studies 
(see {e.g.} \citeauthor{Mitrovica_and_Milne_2002}~\citeyear{Mitrovica_and_Milne_2002}). Some 
features are easily interpreted, such as the broad areas of \sealevel fall across the formerly glaciated 
regions and the surrounding \sealevel rise corresponding to the collapsing forebulges. 
Some other features, such as the equatorial region of \sealevel fall, have a more difficult 
interpretation in terms of ocean siphoning \citep{Mitrovica_and_Milne_2002}. 

The vertical velocity 
($\dot U$) map in Fig.~\ref{fig:udot} clearly shows a strong anti--correlation with $\dot S$, and
in particular it very neatly shows that, in the far field of previously glaciated areas, the 
continents are currently moving up relative to the Earth's center of mass at rates as large as 
$0.5$ mm/yr. This is ultimately the consequence of the differential movements induced by meltwater 
loading acting on the oceans floor. Compared with $\dot S$ and $\dot U$, the rate of sea surface 
variation $\dot N$ in Fig.~\ref{fig:ndot} is characterized by a smoother pattern globally, which 
manifests a relatively large energy content of low--degree harmonics. As seen from
the Earth's center of mass, the sea surface is collapsing everywhere at rates between $0$ and $0.5$ mm/yr,
except in the formerly glaciated regions, where we observe an uplift that broadly follows the uplift
of the solid surface of the Earth in Fig.~\ref{fig:udot}, but with a significantly smaller amplitude. 

In addition to global maps of trends $\dot S$, $\dot U$ and $\dot N$, \selen can similarly produce regional 
analyses, which were not directly accessible in previous release. 
An example is shown in \marginpar{F\ref{fig:sun-med} } Fig.~\ref{fig:sun-med} for the 
Mediterranean region. This figure shows very clearly how GIA is acting across this relatively small, 
mid--latitude basin: the region is subsiding ($\dot U < 0$, middle frame), and though the sea surface 
is collapsing relative to the Earth's center of mass ($\dot N <0$, bottom), in the bulk of the 
Mediterranean relative \sealevel is rising ($\dot S >0$, top). As discussed in e. g. 
\citet{Stocchi_and_Spada_2009}, meltwater loading, described by the second term on the right hand side 
of Eq.~(\ref{load}), is the major cause of subsidence. This is indicated by the shape of the contour 
lines of $\dot U$, which are broadly following the coastlines. 

Tide gauge (TG) records hold a central role in all the assessments of the secular global mean
\sealevel rise  so far published (see \citeauthor{Douglas_1997}~\citeyear{Douglas_1997}, \citeauthor{Spada_and_Galassi_2012}~\citeyear{Spada_and_Galassi_2012} and references therein). Since TGs measure the offset between sea surface and the solid Earth, they 
constitute the experimental realization of the SLE in its basic form, given by Eq.~(\ref{sle-original}). 
Perhaps it is not fully appreciated by the climate change community that the current  
estimates of a global \sealevel rise of $1.8\pm 0.1$ mm/yr since 1880 \citep{Douglas_1997}, recently revised 
to $1.5\pm 0.1$ mm/yr by \citet{Spada_and_Galassi_2012}, tightly 
depend on GIA modeling (hence, to a large extent, on the rheology of the solid Earth). In fact, 
GIA models consistent with RSL observations since the LGM are commonly employed to decontaminate 
TG trends obtained from long records, in order to fully highlight the effect of secular 
climate variations on \sealevel rise (see e. g. \citeauthor{Peltier_2001}~\citeyear{Peltier_2001}). 

Within \selens, the GIA component of \sealevel change at TGs is evaluated by 
\begin{equation}\label{eq:tg}
{r^{gia}_{k}} = \dot{S}(\omega_{k},t_p),
\end{equation}
where $t_{p}$ is present time, the time derivative of \sealevel change is given by Eq.~(\ref{eq:derivative}) and $\omega_{k}=
(\theta_{k},\lambda_{k})$ denote colatitude and longitude of the $k$--th TG. Since the Maxwell relaxation time of the bulk of the mantle 
is of the order of a few kilo--years (see {e.g.}~\citeauthor{Schubert_etal_2004}~ \citeyear{Schubert_etal_2004}), the rate 
(\ref{eq:tg}) can be effectively considered as constant through the period of the instrumental TG record. In \selens, a simple 
analysis can be performed in which predictions of ${r^{gia}_{k}}$ are computed at the TG locations. In the \texttt{TEST} run of 
\selens, we consider the input data collected in file \texttt{DATA/rlr-trends.txt}, pertaining to the $1123$ Revised Local 
Reference (RLR) TG stations listed by the Permanent Service for the Mean Sea Level (PSMSL) as of January 22, 2007 (\selen 
maps for these data are shown in \marginpar{F\ref{fig:tg-distribution}} Fig.~\ref{fig:tg-distribution}). Of course, the user 
can input other data sets, not necessarily containing TG information. Sample outputs of this \selen  analysis are shown in 
\marginpar{T\ref{table:tg1}} Tables \ref{table:tg1} and \marginpar{T\ref{table:tg2}}\ref{table:tg2}. The first shows results 
for TGs with records of at least $100$ years of RLR observations, while the second shows a selection of Mediterranean TGs with 
at least $30$ years of observations. The two tables are showing GIA predictions for the rate of \sealevel change ${r^{gia}_{k}}$, 
but also for the other fundamental geodetic quantities $\dot U$ and $\dot N$ (a new feature with respect to \selen 1.0). All data 
and plots produced by \selen for the TG analyses illustrated above are found in folder \texttt{depot-TEST/tgauges}.

As a final example of a possible geodetic application available within \selens, we consider the time--variations
of the gravity potential harmonic coefficients. This analysis can be useful, for example, to evaluate the effects of GIA 
on the gravity variations observed by the NASA/DLR Gravity Recovery and Climate Experiment (GRACE) satellites (see the overview of \citeauthor{Tapley_etal_2004}~\citeyear{Tapley_etal_2004}). According to \selen conventions, the GIA--induced total 
variation of the gravity field potential, evaluated at the Earth's surface is:  
\begin{equation}\label{eq:sh-expansion-selen}
\Phi(a,\omega) = \frac{\Gamma M_{e}}{a} \sum_{\ell=2}^{\ell_{max}} \sum_{m=-\ell}^{\ell} \mathbf{\Phi}_{\ell m} {\cal Y}_{\ell m}(\omega),
\end{equation}
where $\Gamma$ is Newton's constant, $M_{e}$ and $a$ are the mass and the average radius of the Earth, 
$\ell_{max}$ is the maximum harmonic degree of the analysis (for run \texttt{TEST}, $\ell_{max}=128$), $\mathbf{\Phi}_{\ell m}$ 
are the gravity coefficients in complex form (these obey $\mathbf{\Phi}_{\ell -m}$=$\mathbf{\Phi}_{\ell m}^\ast$)
and ${\cal Y}_{\ell m}(\omega)$ are the complex $4\pi$--normalized SH functions of harmonic degree $\ell$ and order $m$:
\begin{equation}\label{eq:ylm}
{\cal Y}_{\ell m}(\omega) = 
\sqrt{(2\ell +1)\frac{(\ell -m)!}{(\ell +m)!}}
P_{\ell m} (\cos\theta) \textrm{e}^{im\lambda}, 
\end{equation}
where $P_{\ell m}(\cos\theta)$ are the associated Legendre functions.
Note that in Eq.~(\ref{eq:sh-expansion-selen}), terms of harmonic degree $\ell=0$ are not included since the total mass
of the Earth is conserved, and those of degree $\ell=1$ vanish since the origin of the reference frames is assumed
to coincide with the Earth's center of mass. It is sometimes more convenient to transform Eq.~
(\ref{eq:sh-expansion-selen}) into an equivalent expansion involving real coefficients. For instance, 
according to the conventions usually adopted in gravimetry (including GRACE), we can equivalently write
\begin{equation}\label{eq:sh-expansion-grace}
\Phi(a,\omega) = \frac{\Gamma M_{e}}{a} \sum_{\ell=2}^{\ell_{max}} \sum_{m=0}^{\ell}\left( {\bar c}_{\ell m} \cos m\lambda + {\bar s}_{\ell m} 
\sin m\lambda \right)\bar{P}_{\ell m}(\cos \theta),
\end{equation}
where $\bar{P}_{\ell m}(\cos \theta)$ denotes the ``fully normalized'' associated Legendre functions, 
\textit{not} including the Condon--Shortley phase $(-1)^{m}$. After some straightforward algebra, for the
variations of the Stokes coefficients one obtains 
$\left( {\bar c}_{\ell m}, {\bar s}_{\ell m}\right)$ = $(-1)^{m}$ $\sqrt{2-\delta_{0m}}~\mathbf{\Phi}^\ast_{\ell m}$.
Program \texttt{stokes.f90} provides, degree--by--degree, the time derivatives ${\dot{\bar{c}}}_{\ell m}$ 
and ${\dot{\bar{s}}}_{\ell m}$, which can be directly compared with trends obtained from monthly GRACE observations 
(derivatives are numerically evaluated following Eq.~\ref{eq:derivative}) in order to assess the GIA effects. 
\marginpar{F\ref{fig:stokes}} Fig.~\ref{fig:stokes} shows the result obtained for run \texttt{TEST}, for harmonics up 
to degree $\ell_{max}=9$. A similar analysis has been 
performed recently by \citet{Sorensen_2010}, using various global ice models from the literature (and available
in the ice models directory \texttt{ICE-MODELS} of \selens). 

\section{Conclusions}\label{sec:conclusions}

We have described the basic features and the essential theory background of \selens, 
an open source Fortran $90$ program that solves numerically the SLE. Version 2.9, which constitutes a significant 
improvement of previous version \selen 1.0, has been developed to respond to the requests of colleagues working on various aspects of GIA, who did not have access to a SLE solver, and for teaching purposes. 
\selens, which is the numerical implementation of the SLE theory presented by \citet{Farrell_and_Clark_1976}, 
can be used to efficiently simulate the spatiotemporal variations of several geophysical and geodetic quantities involved in the GIA process. These include relaxation spectra and LDCs computed using user--supplied rheological profiles, the spatial distribution of the continental ice sheets according to several built--in GIA models, predictions of relative \sealevel variations at specific sites from which observations are available since the LGM, global and regional maps of present-day rates of variations of GIA--induced
\sealevel change and displacements, local predictions for geodetic variations (including vertical and horizontal movements), and time--variations of harmonic coefficients of the gravity field in response to GIA. 
Refined implementations of \selen 1.0 have been recently employed in various geodetic contexts, ranging from
the interpretation of GPS data on a regional scale in Greenland \citep{Nielsen_etal_2012},  
to the correction of GRACE observations aimed at estimating the mass variation 
in the Mediterranean and in the Black Sea \citep{Fenoglio-Marc_etal_2012}, and to the study of the 
effects of GIA in West Antarctica \citep{Groh_etal_2012} and Greenland \citep{Ewert_etal_2012,Spada_etal_2012}.

\selen is written in standard Fortran 90 and takes advantage of an OpenMP multi--threaded parallelism. It can be 
installed and compiled on Mac OS, Linux and Windows platforms. Starting from the examples illustrated in this manuscript, the 
users can customize \selen in order to solve \textit{ad--hoc} GIA problems or to include new functionalities 
within the code. According to our experience with the students of the $2$nd Training School on Glacial Isostatic 
Adjustment (GIA) Modeling, where \selen has been employed for a tutorial introduction to the physics and the 
phenomenology of GIA, the program may constitute a valuable instrument for scientists at their first approach to 
this problem. New features such the simulation of the horizontal migration of shorelines in response to 
\sealevel change and the realization of the feedback between \sealevel variations and rotational fluctuations, 
are under way and will be available with the upcoming version of \selen (v. 3.0). 

\section*{Acknowledgments}
We thank all the students and the participants to $2$nd Training School on Glacial Isostatic Adjustment Modeling, held in G\"avle (Sweden) in June $2011$, and all the \selen users for the numerous feedbacks that have contributed to improve the program. Valentina Barletta is acknowledged for providing benchmark test computations and for advice. All computations involving spherical harmonic functions are performed using routines from the \texttt{SHTOOLS} package by Mark Wieczorek (see \textit{http://www.ipgp.fr/wieczor/SHTOOLS/SHTOOLS. html}). We thank Max Tegmark for making available the Fortran code for the icosahedral pixelization of the celestial sphere, which is now incorporated in \selens.  All the figures have been drawn using the GMT public domain software \citep{Wessel_and_Smith_1998}.  \selen is available upon request to the authors. The numerical computations have been partly performed thanks to a CINECA award under the ISCRA (Italian SuperComputing Resource Allocation) initiative, for the availability of high performance computing resources and support through the Class C Projects contract n. HP10CS9J50. This work has been supported by COST Action ES0701 ``Improved Constraints on Models of Glacial Isostatic Adjustment''.


\clearpage
\begin{table}
\centering
\begin{tabular}{ll}
\hline
\texttt{config.dat}                    & \selen configuration file \\
\texttt{makeselen.sh}                    & Bash script that executes \selen \\
\\
\underline{\selen Input data:} & \\
\texttt{wdir/DATA}& Various input data for RSL analyses and more \\
\texttt{wdir/ICE-MODELS}& Ice models data \\
\texttt{wdir/VSC}& Collection of viscosity profiles \\
\\
\underline{Fortran $90$ units:} & \\
\texttt{config.f90} &  Interpreter of the configuration file  \\
\texttt{esl.f90} & Equivalent Sea Level \\
\texttt{geo.f90} & Time variations of geodetic quantities \\
\texttt{gmaps.f90} & Synthesis of geodetic quantities on global maps \\
\texttt{harmonics.f90} & Include file with Various SH tools and utilities \\
\texttt{ms.f90} & GMT multi--segment files from ice data \\
\texttt{of\_dv.f90}& Degree variance of the Ocean Function \\
\texttt{px.f90}& Pixelization tools (including the Tegmark algorithm)\\
\texttt{px\_rec.f90}& Reorganizes the pixelization data \\
\texttt{rec\_ice.f90}& SH reconstruction of the ice thickness \\
\texttt{rec\_of.f90}& SH reconstruction of the Ocean Function  \\
\texttt{rmaps.f90}& Synthesis of geodetic quantities on regional maps \\
\texttt{rsl.f90}& Relative Sea Level curves \\
\texttt{rsl\_zones.f90}& Geometry of the Relative Sea Level ``Clark's zones''  \\
\texttt{rslc.f90}& Relative Sealevel Contour lines for regional analyses \\
\texttt{sh.f90}& SHs at the grid pixels \\
\texttt{sh\_of.f90}& SH coefficients for the Ocean Function \\
\texttt{sh\_rsl.f90}& SHs at the RSL sites \\
\texttt{sh\_rslc.f90}& SHs for regional analysis and RSL contours  \\
\texttt{sh\_tgauges.f90}& SHs at the TG sites  \\
\texttt{shape\_factors.f90}& ``Shape factors'' for the ice elements \\
\texttt{shice.f90}& SH decomposition of the ice model \\
\texttt{shtools.f90}& An SHTOOLS interface for the SH analysis \\
\texttt{sle.f90}& The SLE solver \\
\texttt{stokes.f90}& Variations of the Stokes coefficients of the gravity field \\
\texttt{tgauges.f90}& Present--day rate of \sealevel change at the TG sites \\
\texttt{tb.F90}& The TABOO code \\
\texttt{wnw.f90} &  Numerical test for the SH orthogonality \\
\hline
\label{table:some-files}
\end{tabular}
\caption{Some of the files contained in the \selen package. Here \texttt{wdir} indicates the working directory of \selens.}
\end{table}
\clearpage

\begin{table}
\centering
\begin{tabular}{ll}
\hline
\\
\texttt{Folder:}                   &  \texttt{content:}  \\
\\
\texttt{ICE5G}                     & Data about the ice model\\
\texttt{~~~ICE5G/esl}                     & ESL \\
\texttt{~~~ICE5G/original}                     & Ice thickness data \\
\texttt{~~~ICE5G/reconstructed}                     & SH reconstruction of ice thickness data\\
\texttt{~~~ICE5G/sh}                     & SH coefficients of the ice model \\
\texttt{Love-Numbers-by-TABOO}            & Love numbers data\\
\texttt{TABOO}   & TABOO input files \\
\texttt{bin}   & An empty folder\\
\texttt{geod}   & Predictions at geodetic sites\\
\texttt{~~~geod/3dmaps}   & Regional maps (in progress) \\
\texttt{~~~geod/sites}   & Geodetic predictions at specific sites \\
\texttt{gmaps}   & Global geodetic rates (data and plots) \\
\texttt{log}   & Log files of \selen and TABOO \\
& (with summary of Earth parameters) \\
\texttt{of}   & Ocean function data and plots \\
\texttt{~~~of/degree\_variance}   & Ocean function degree variance \\
\texttt{px}   & Various pixelization data and plots \\
\texttt{rmaps}   & Regional geodetic rates (data and maps) \\
\texttt{~~~rmaps/Antarctica}   & rates for Antarctica  \\
\texttt{~~~\ldots} & \texttt{\ldots} \\
\texttt{~~~rmaps/North\_America}   & rates for North America  \\
\texttt{rsl}   & Relative Sea Level (RSL) data folder \\
\texttt{~~~rsl-contours} & RSL contour plot \\
\texttt{~~~rsl-curves} & RSL curves at specific sites \\
\texttt{~~~rsl-misfit}  & Misfit between RSL data and predictions \\
\texttt{~~~rsl-scplot} & Scatterplot of RSL data \\
\texttt{~~~rsl-sites} & Data and plots regarding RSL sites \\
\texttt{~~~rsl-table} & Summary table of RSL data and predictions \\
\texttt{~~~rsl-zones} & RSL zones \\
\texttt{stokes}   &  Stokes coefficients data \\
\texttt{tgauges}   & Tide gauges (TGs) \\
\texttt{~~~tgauges-predictions}  & Predictions at TGs   \\
\texttt{~~~tgauges-scplots}  & TG data scatterplot \\
\texttt{~~~tgauges-sites} & Maps of TG sites\\
\texttt{wnw}   & ``Window test'' for the ocean function \\
\hline
\label{table:folders}
\end{tabular}
\caption{Organization of the \selen outputs in directory \texttt{depot-TEST}. \selen outputs 
consist of various plain text, postscript, and pdf files. }
\end{table}

\begin{figure}
\begin{center}
\noindent\includegraphics[width=1\textwidth, angle=0]{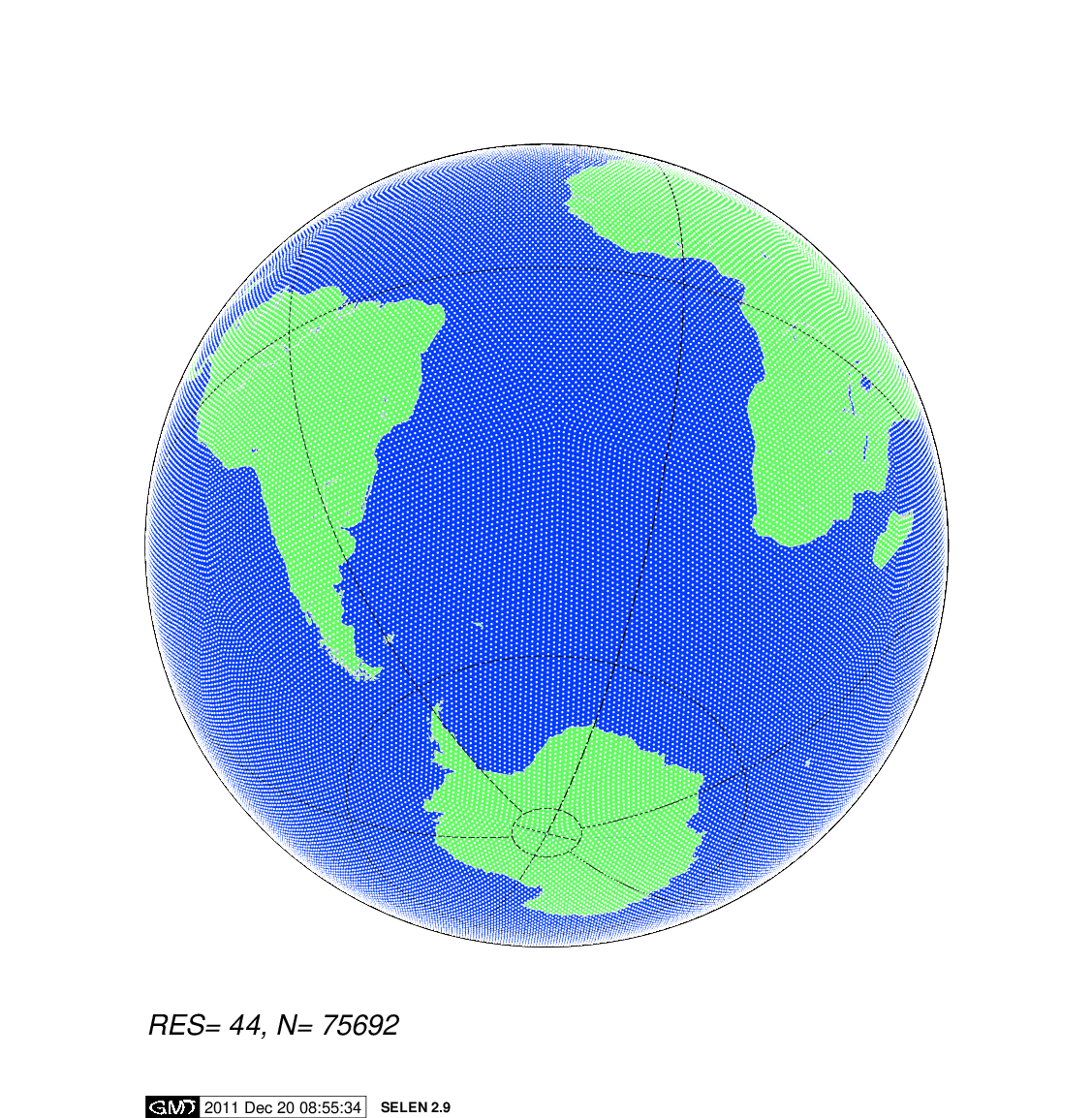}
\end{center}
\caption{Pixelization of the sphere following the icosahedron--based method proposed by \citet{Tegmark_1996} for astrophysical applications, 
implemented in \selens. This geometrical tool provides a natural set of Gauss points on the sphere and allows for a straightforward computation of integrals involving SH functions. The number of pixels in the grid is $N_{p}=40{R}({R}+1)+12$, where $R$ is the resolution parameter.
Here ${R}=44$ (the corresponding grid spacing is $\approx 45$ km). The constraint $N_{p} \ge \ell^{2}_{max}/3$, which ensures an optimal integration on the sphere \citep{Tegmark_1996} is largely met (in the \texttt{TEST} run, $\ell_{max}=128$). Data about the grid, including spherical coordinates of ``wet'' (oceanic, blue) and  ``dry'' (continental, green) pixels and postscript figures are found in folder 
\texttt{depot-TEST/px}. In \selens, wet pixels are separated by dry pixels using the GMT program \texttt{gmtselect} 
\citep{Wessel_and_Smith_1998}. By default, \selen employs the full resolution coastlines of GMT (\texttt{-Df}), and dry (wet) 
pixels are selected using option \texttt{-Ns/k/s/k/s} (\texttt{-Nk/s/k/s/k}) of \texttt{gmtselect}.}
\label{fig:pixelization}  
\end{figure}
\clearpage  

\newpage
\begin{table}
\centering
\begin{tabular}{lccccc}
\hline
Layer                           & Radius                        &Density                        &Shear modulus          &Viscosity                              &       Gravity\\
                                       & (km)                  &(kg m$^{-3}$)          &($\times 10^{11}$ Pa)  &($\times 10^{21}$Pa s) &(m s$^{-2}$)\\
\hline
Lithosphere                     & $6281$--$6371$        &$4120$                 &$0.73$                 &$\infty$                               &$9.707$\\
Shallow upper mantle    & $5951$--$6281$        &$4120$                 &$0.95$                 &$0.5$                          &$9.672$\\
Transition zone                 & $5701$--$5951$        &$4220$                 &$1.10$                 &$0.5$                          &$9.571$\\
Lower mantle                    & $3480$--$5701$        &$4508$                 &$2.00$                 &$2.7$                          &$9.505$\\
Core                            & $0$--$3480$   & $10925$               &$0.00$                         &$0$                            &$10.622$\\
\hline
\label{table:vm2}
\end{tabular}
\caption{Model parameters for model VM2a, employed for the \texttt{TEST} run of \selens. The whole library of models available
within \selen is accessible in the Fortran unit \texttt{tb.F90}.}
\end{table}
\clearpage

\begin{figure}
\begin{center}
\noindent\includegraphics[width=0.9\textwidth, angle=-90]{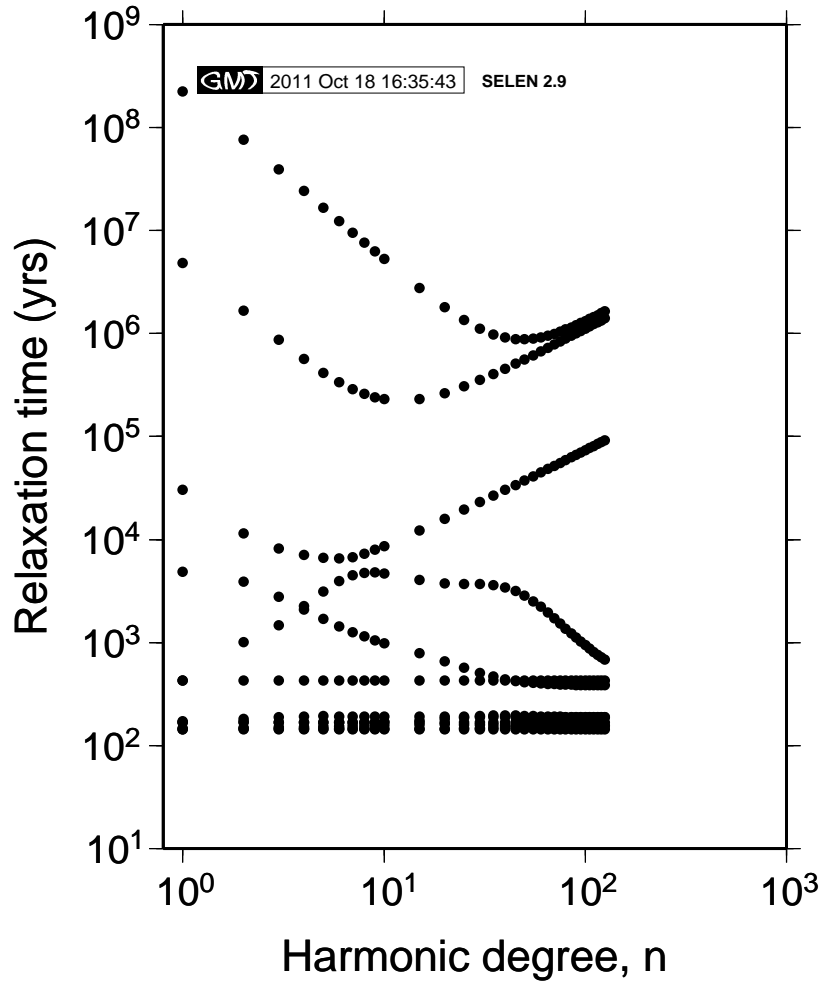}
\caption{Isostatic relaxation spectrum for model VM2a (see Table \ref{table:vm2}), showing the relaxation times as a function of 
harmonic degree $\ell = n$ in the range $1\le \ell \le 128$. The physical meaning 
of the spectrum is discussed by \citet{Spada-2003}.}
\label{fig:spectrum}
\end{center}
\end{figure}
\clearpage

\begin{figure}
\begin{center}
\noindent\includegraphics[width=0.6\textwidth, angle=-90]{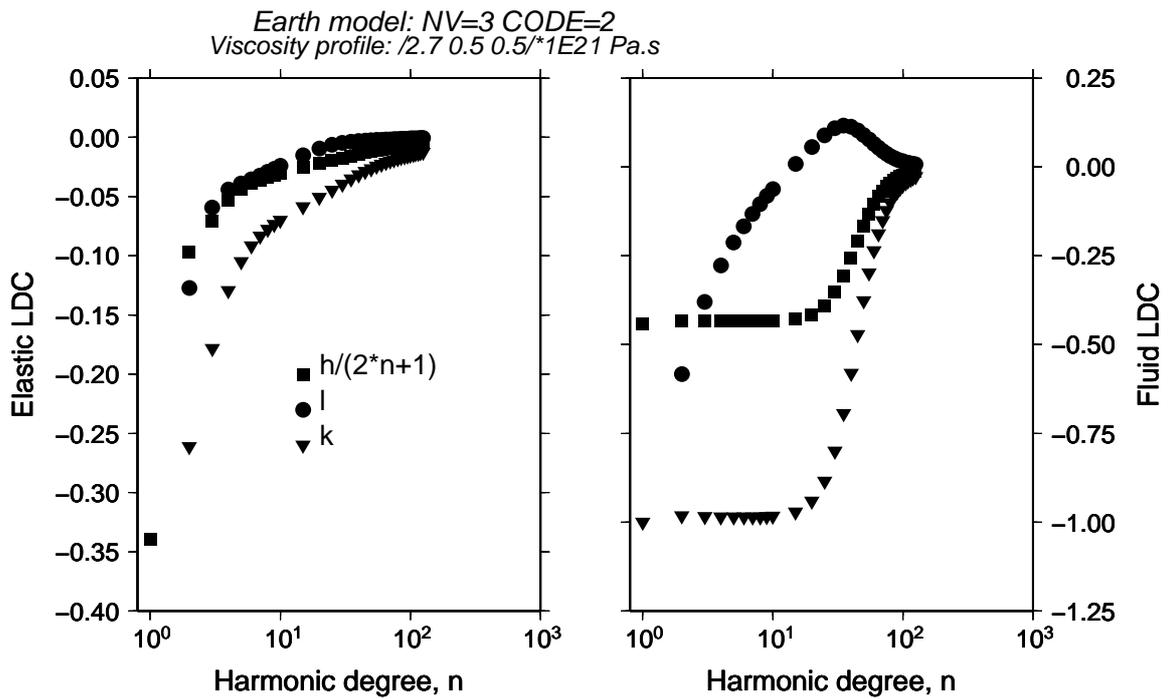}
\caption{Elastic (left) and fluid (right) values of the LDCs 
$h$ (associated to vertical displacement), $l$ (horizontal displacement) and $k$ (incremental gravitational potential) 
for model VM2a (see Table~\ref{table:vm2}), as a function of harmonic degree $\ell$. 
Note that $h$ is normalized by $(2\ell +1)$. For the definition of the LDCs
see e.g.~\citet{Spada-2003}.}
\label{fig:ldcs}
\end{center}
\end{figure}
\clearpage

\begin{figure}
\begin{center}
\includegraphics[width=0.9\textwidth, angle=-90]{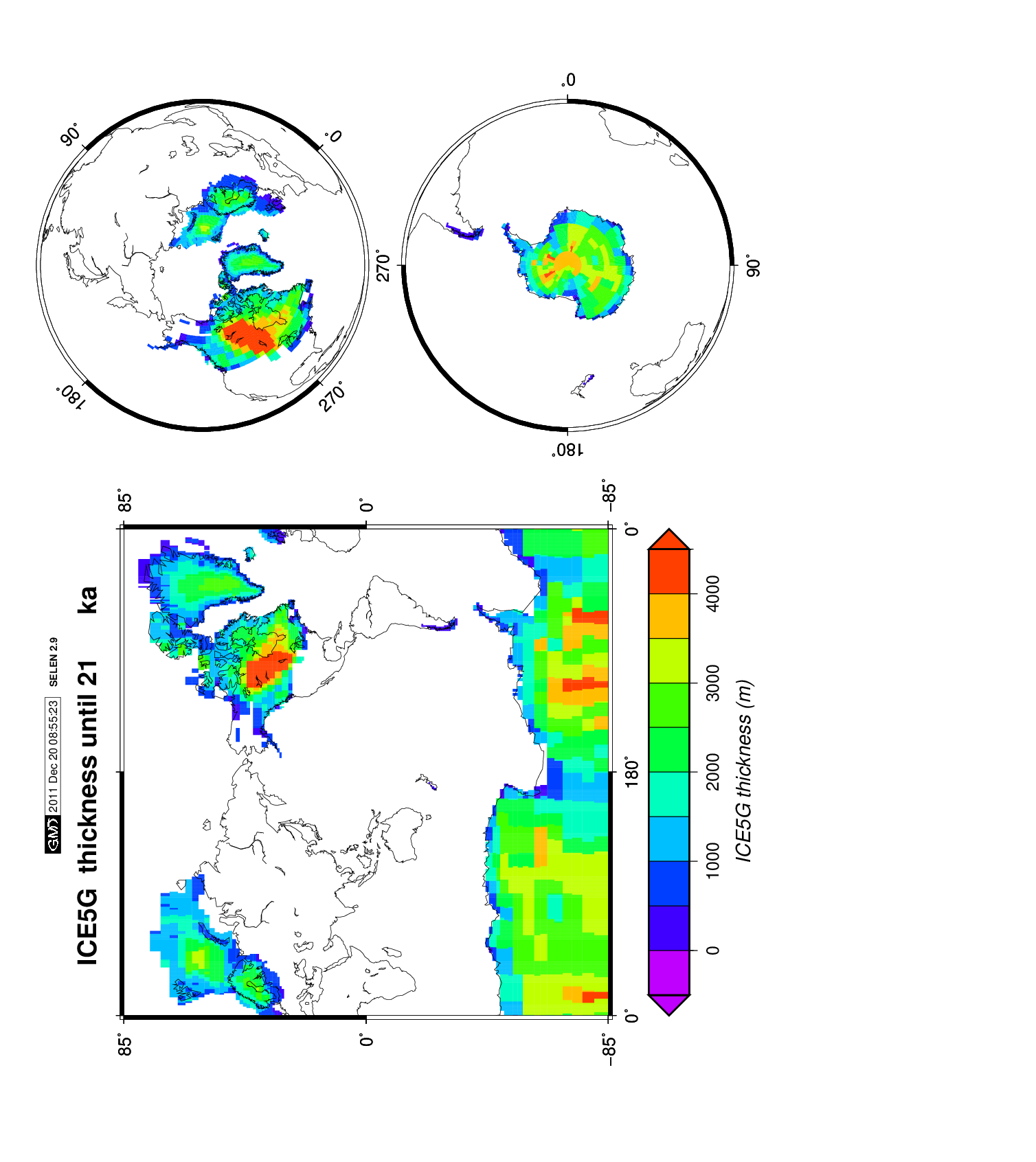}\vspace{-10em}
\includegraphics[width=0.9\textwidth, angle=-90]{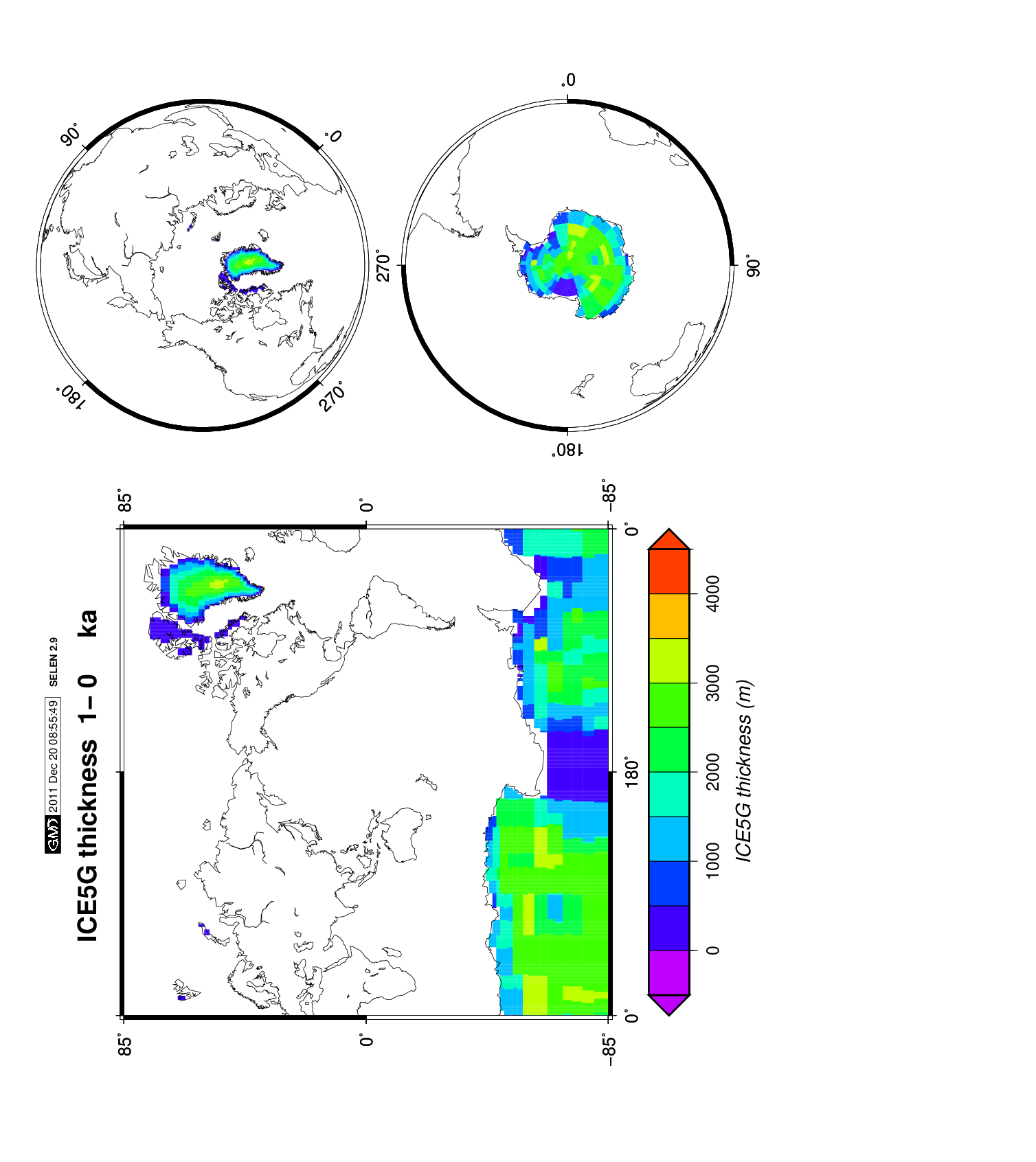}\vspace{-10em}
\caption{Ice thickness $T(\omega,t)$ at the LGM ($21$ kyrs ago, top) and at during the most recent 
time increment, between $1$ kyr BP and present time (bottom), according to model ICE--5G \citep{Peltier_2004}. 
The time--history of the Equivalent Sea Level for this model is shown in Fig. \ref{fig:eustatic}. Maps
for all the other time steps between LGM and present 
are available in folder \texttt{depot-TEST/ICE5G/original}.}
\label{fig:ice5g}
\end{center}
\end{figure}
\clearpage  

\begin{figure}
\begin{center}
\noindent\includegraphics[width=0.6\textwidth, angle=-90]{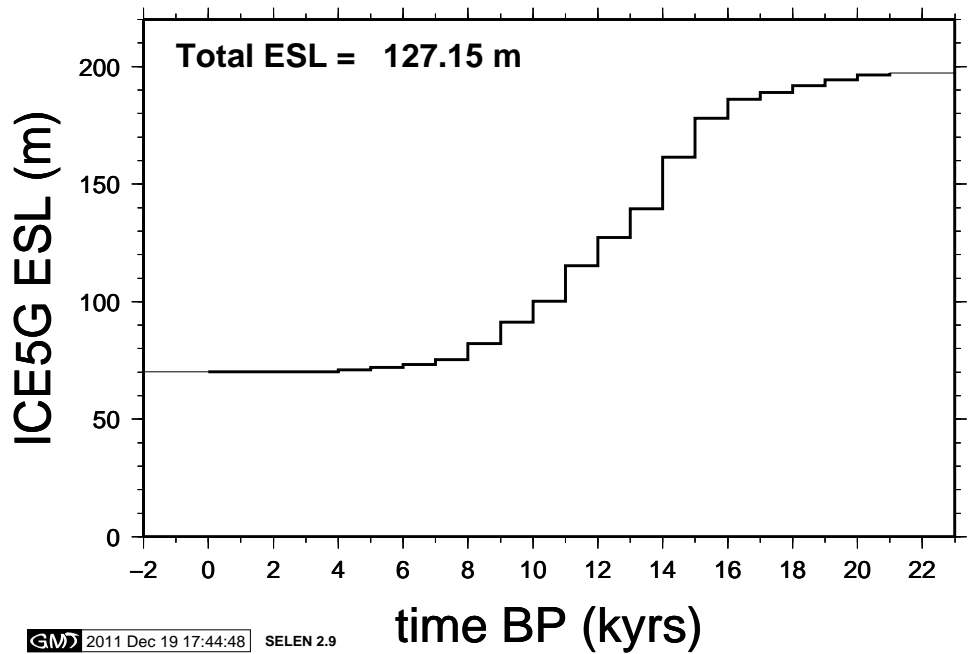}
\caption{Equivalent Sea Level for model ICE--5G \citep{Peltier_2004}. At a given time $t$ before present, 
$\textrm{ESL}(t)= (\rho_i/\rho_w)(V_{i}(t)-V_{i}(t_{p}))/A_{o})$, where  $V_{i}(t)$ is the ice volume, 
$V_{i}(t_{p})$ is present day volume, and $A_{o}$ is the area of the oceans surface. Hence, 
according to Eq.~(\ref{eustatic}), the plot of ESL mirrors that of $S^{E}$. The total
ESL variation ($\sim 127$ m) represents the difference between ESL at the LGM and the present day value.}
\label{fig:eustatic}
\end{center}
\end{figure}
\clearpage 

\begin{figure}
\begin{center}
\noindent\includegraphics[width=0.80\textwidth, angle=-90]{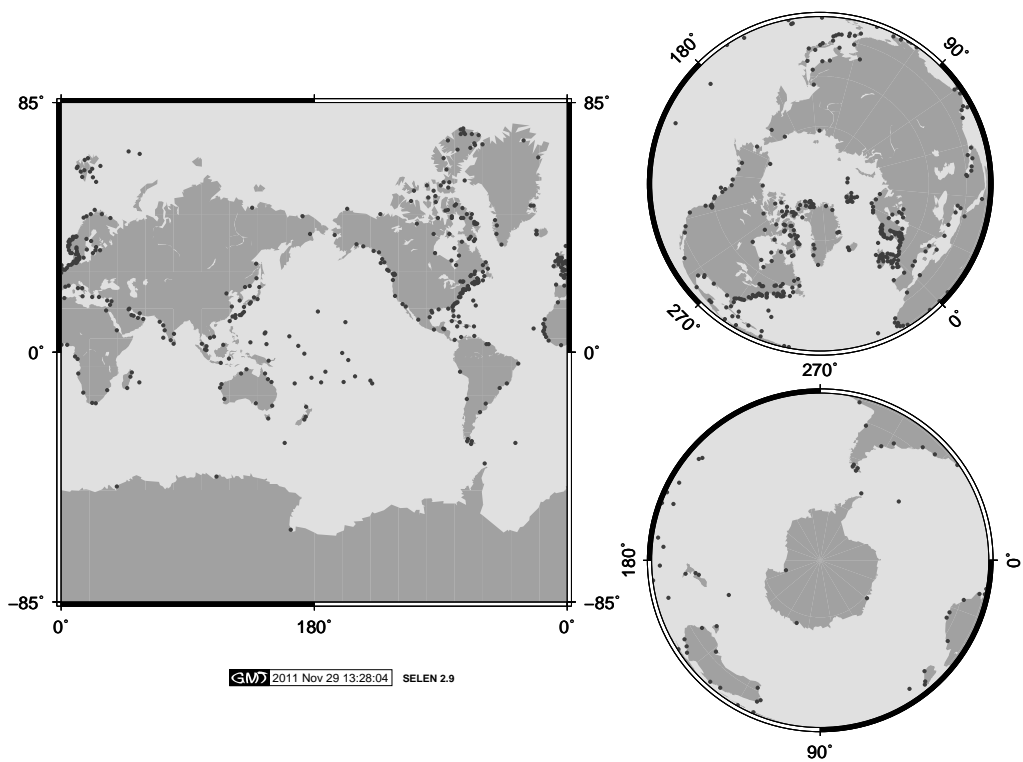}
\caption{Geographical distribution of the $392$ sites in file \texttt{sealevel.dat}, from which 
information about the history of RSL during the last $\sim 15,000$ years is available. This plot 
and more data about these sites are available from \texttt{depot-TEST/rsl/rsl-sites}.}
\label{fig:rsl-sites}
\end{center}
\end{figure}
\clearpage

\begin{figure}
\begin{center}
\noindent\includegraphics[width=0.8\textwidth, angle=0]{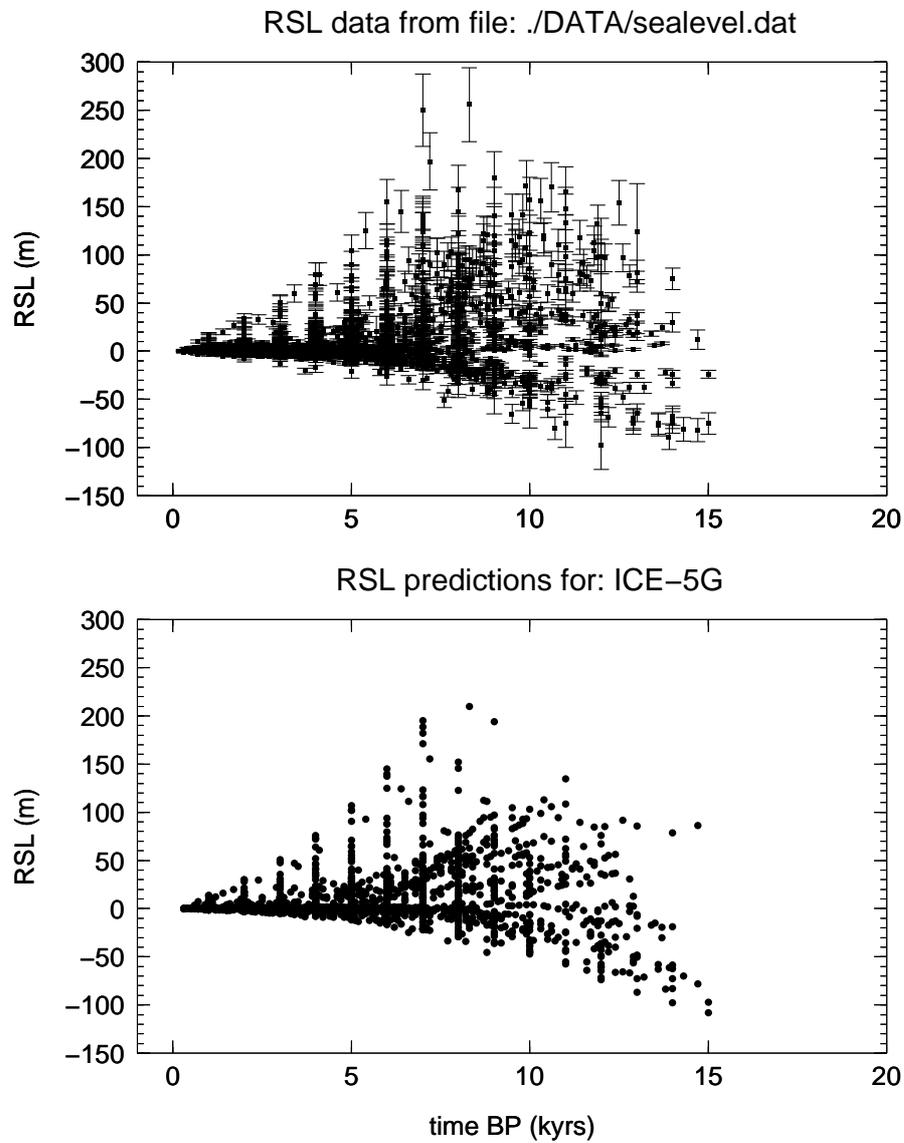}
\caption{Scatterplot showing RSL observations (top) from sites of the compilation of 
\citet{Tushingham_and_Peltier_1992,Tushingham_and_Peltier_1993}. 
RSL predictions, obtained solving the SLE in our \texttt{TEST} run of \selens, 
are shown in the bottom frame.}
\label{fig:rsl-scatterplot}
\end{center}
\end{figure}
\clearpage

\begin{figure}
\begin{center}
\noindent\includegraphics[width=0.9\textwidth,angle=0]{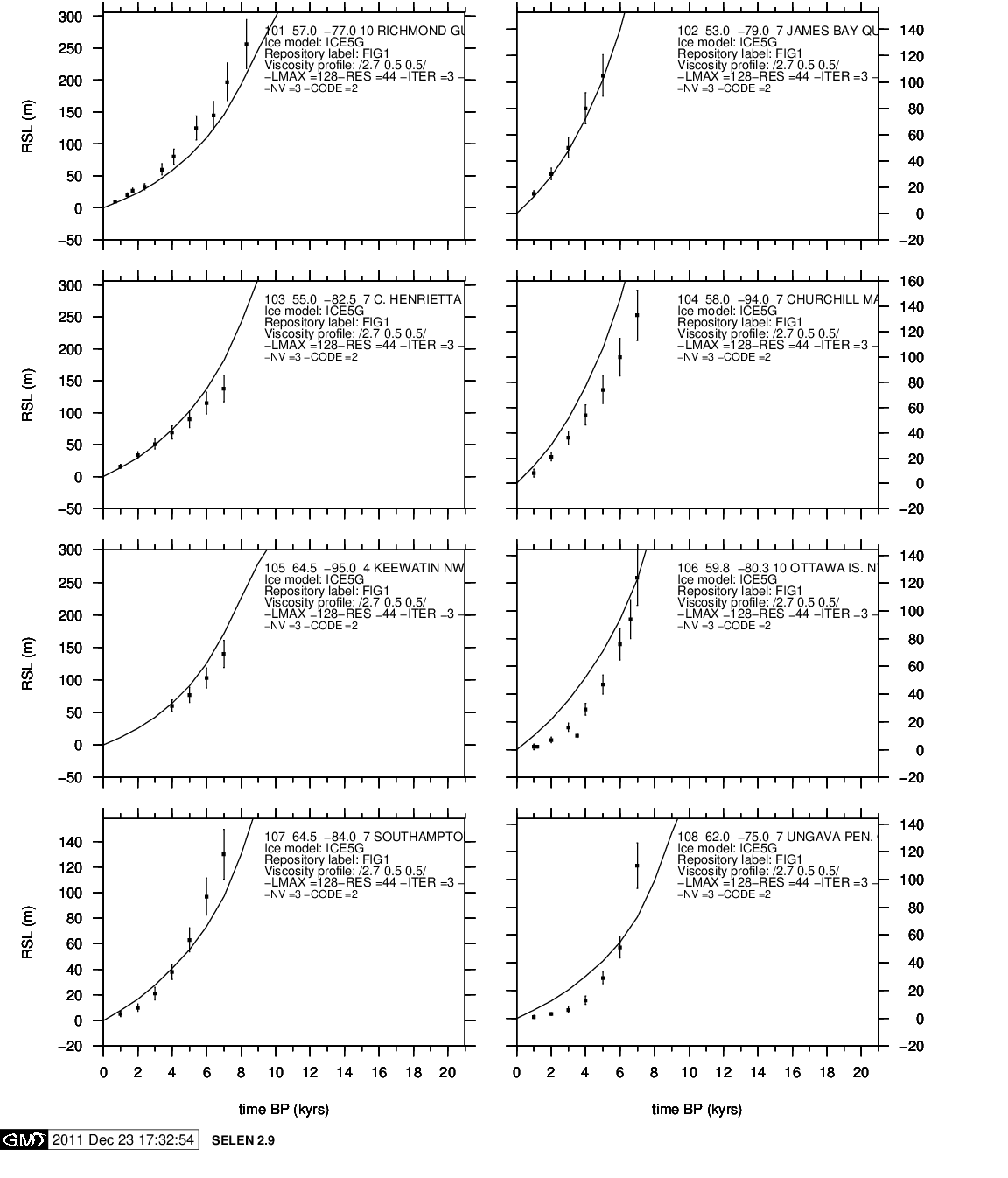}
\caption{RSL observations (with error bars) pertaining to the eight sites of Hudson bay
in file \texttt{sealevel.dat}, compared with \selen predictions (solid curves). Basic parameters 
for the \texttt{TEST} run are summarized in each frame. Postscript and PDF figures are located  
\texttt{depot-TEST/rsl/rsl-curves/ps} and \texttt{depot-TEST/rsl/rsl-curves/pdf}, respectively.}
\label{fig:rsl-hbay}
\end{center}
\end{figure}
\clearpage

\begin{figure}
\begin{center}
\noindent\includegraphics[width=0.9\textwidth]{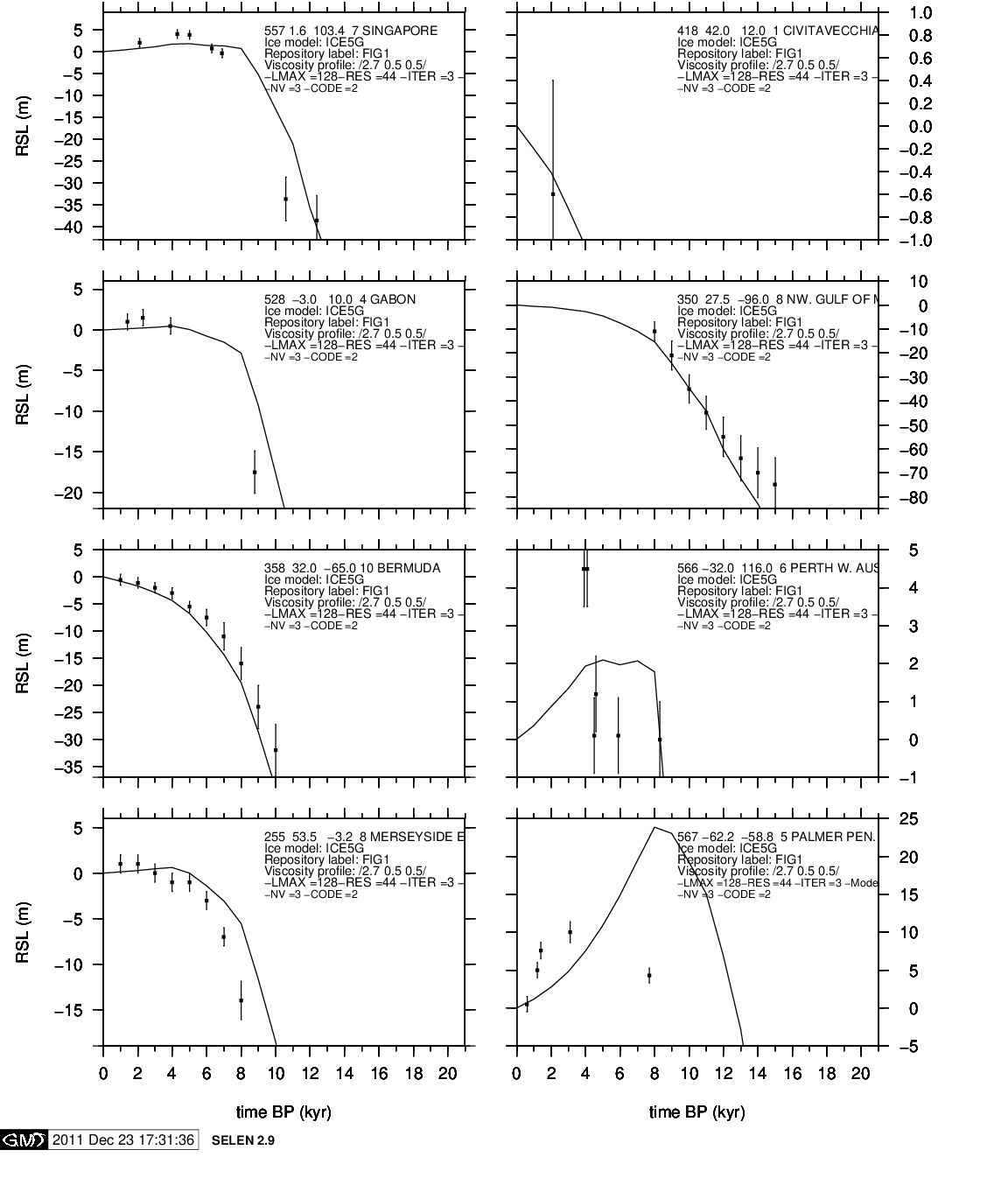}
\caption{RSL curves at eight miscellanea sites located in the far field of the former ice sheets:
Singapore, Civitavecchia (Italy), Gabon, NW Gulf of Mexico, Bermuda, Perth W. Australia, 
Merseyside (England), and Palmer peninsula (Antarctica). All the RSL predictions for the
\texttt{TEST} run are found in \texttt{depot-TEST/rsl/rsl-curves}.}
\label{fig:rsl-var}
\end{center}
\end{figure}
\clearpage

\begin{figure}
\begin{center}
\noindent\includegraphics[width=0.8\textwidth, angle=0]{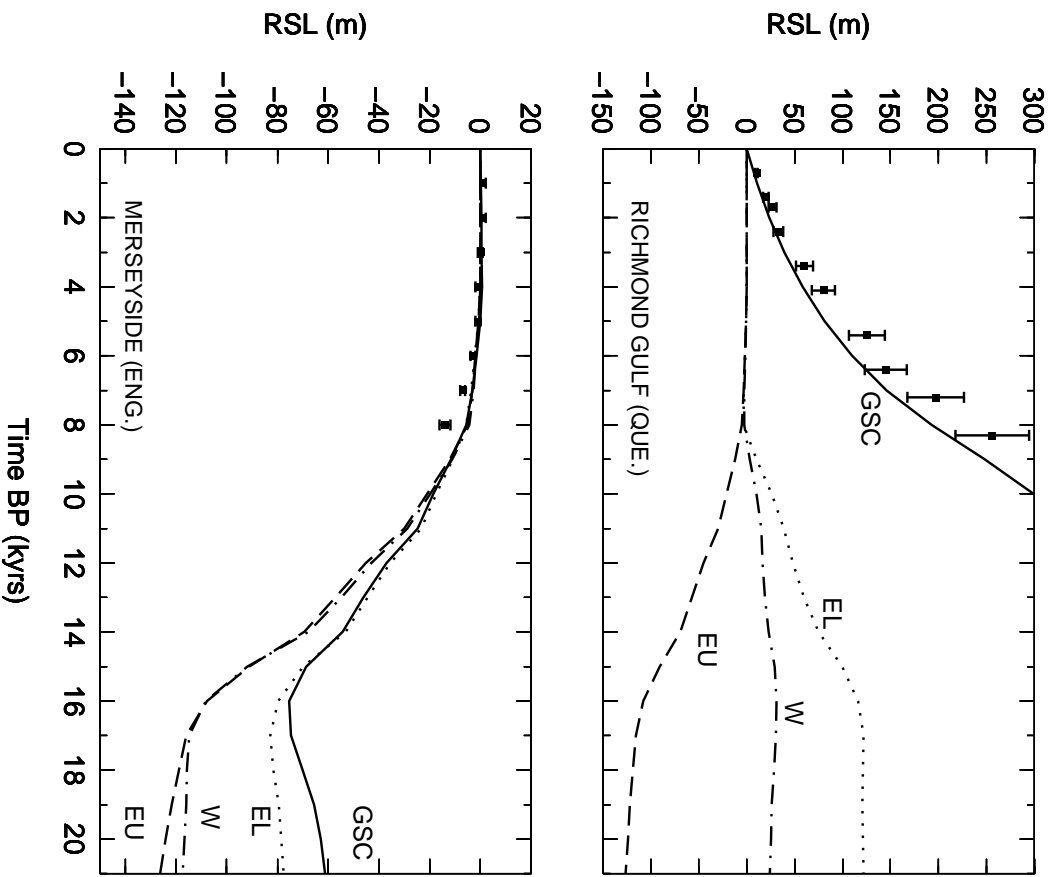}
\caption{Comparison of the outcomes of various \selen runs, in which the SLE is solved using
various approximations (GSC: gravitationally self--consistent, EL: elastic, W: Woodward, EU: Eustatic). 
Top and bottom frames show
results for Richmond Gulf (Hudson bay) and Merseyside (England), respectively. }
\label{fig:rsl-sleapp}
\end{center}
\end{figure}
\clearpage

\begin{figure}
\begin{center}
\noindent\includegraphics[width=1.2\textwidth, angle=-90]{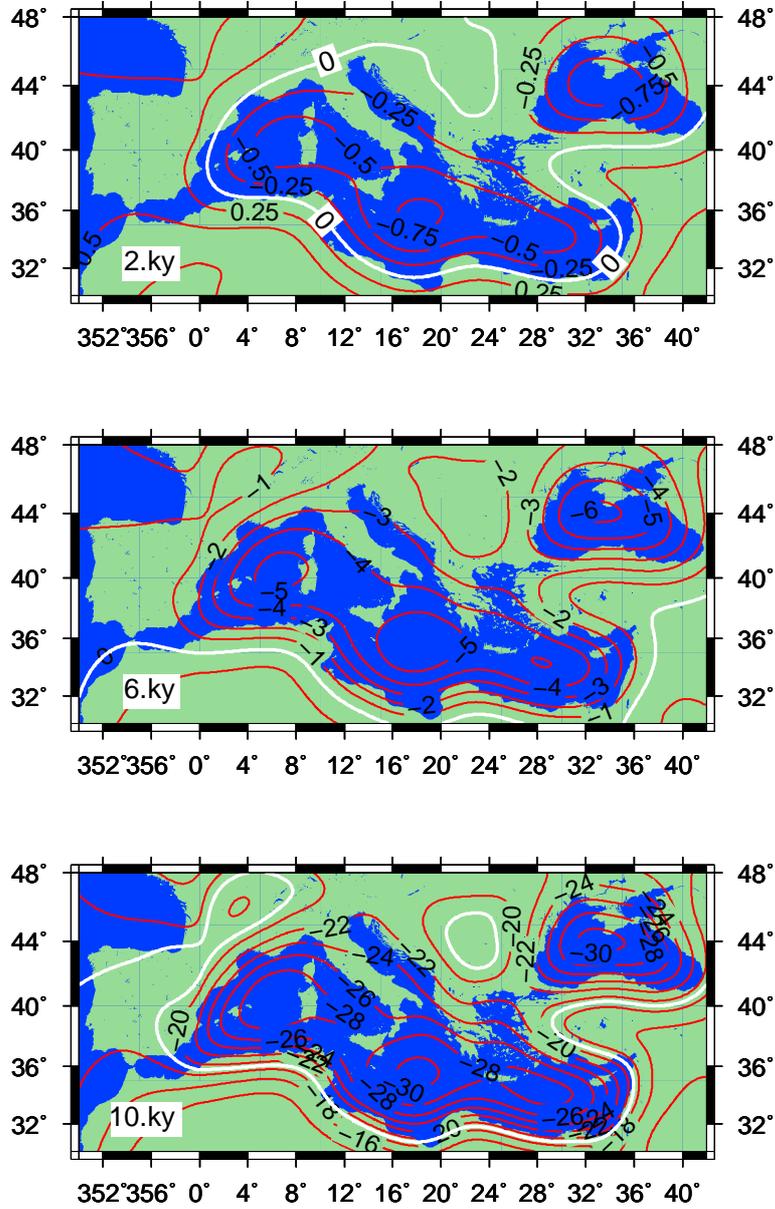}
\caption{Example of RSL contour plot for the Mediterranean region, at times $2$, $6$, and $10$ kyrs 
BP. These correspond to three different configurations of file \texttt{DATA/rsl-region.dat}, all
based on the \selen parameters employed in the \texttt{TEST} run (ice model ICE--5G and rheological
parameters in Table \ref{table:vm2}).}
\label{fig:med-rslc}
\end{center}
\end{figure}
\clearpage

\begin{figure}
\vspace{2cm}
\hspace{3cm}
\includegraphics[width=1.1\textwidth, angle=0]{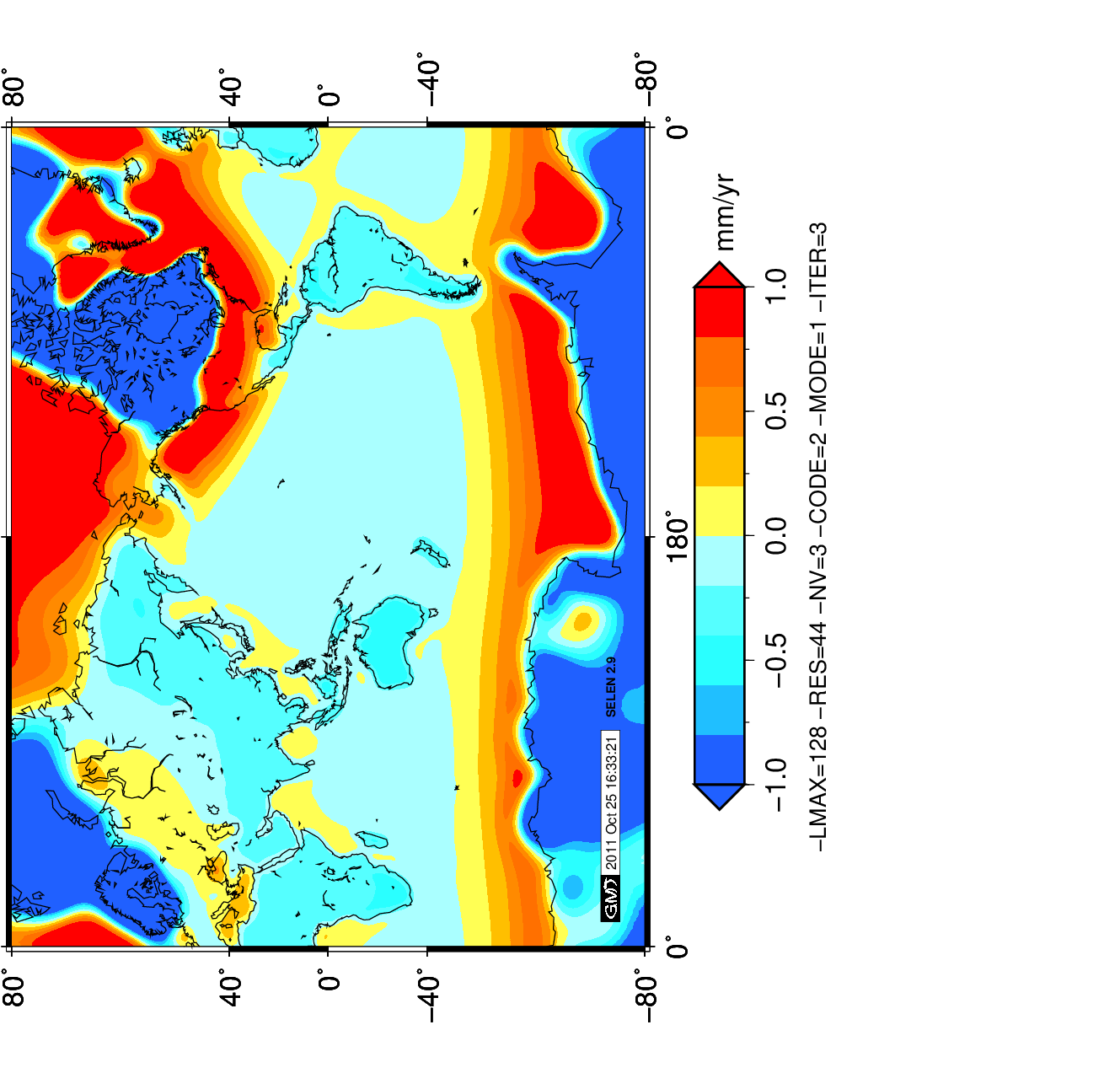}
\caption{Global map showing the present--day rate of \sealevel change associated with GIA ($\dot S$)
for our \texttt{TEST} run. Data and plots for this analysis are found in folder \texttt{depot-TEST/gmaps}.
The $\dot S$ values varies in the range $[-17.0/3.7]$ mm/yr.}
\label{fig:sdot}
\end{figure}
\clearpage

\begin{figure}
\vspace{2cm}
\hspace{3cm}
\includegraphics[width=1.1\textwidth, angle=0]{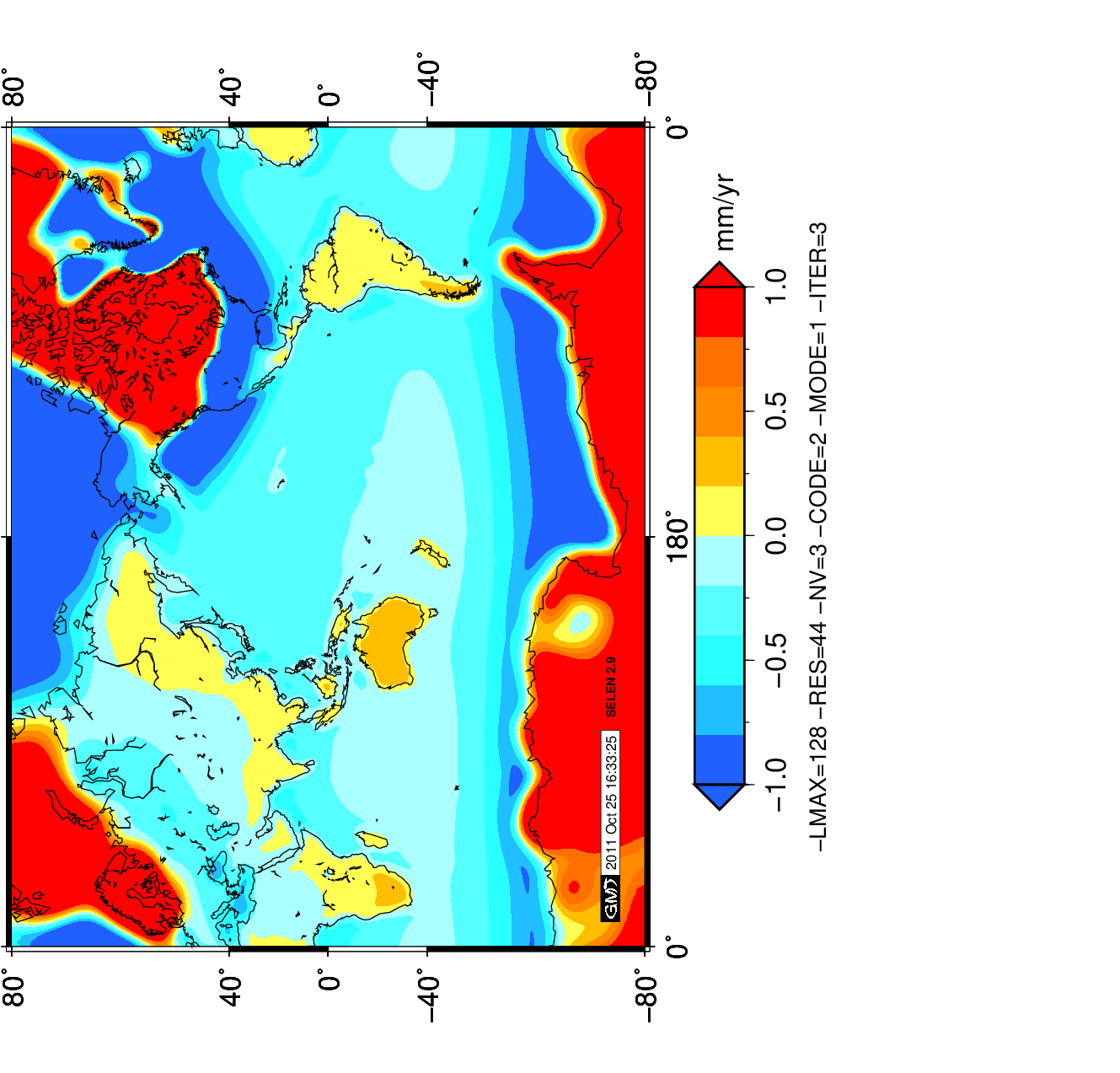}
\caption{Global map of GIA--induced vertical velocity ($\dot S$) for our \texttt{TEST} run. Subsiding
and uplifting areas are shown by blue and red hues, respectively. In this map, the $\dot U$ values vary 
in the range $[-3.5/19.2]$ mm/yr.}
\label{fig:udot}
\end{figure}
\clearpage

\begin{figure}
\vspace{2cm}
\hspace{3cm}
\includegraphics[width=1.1\textwidth, angle=0]{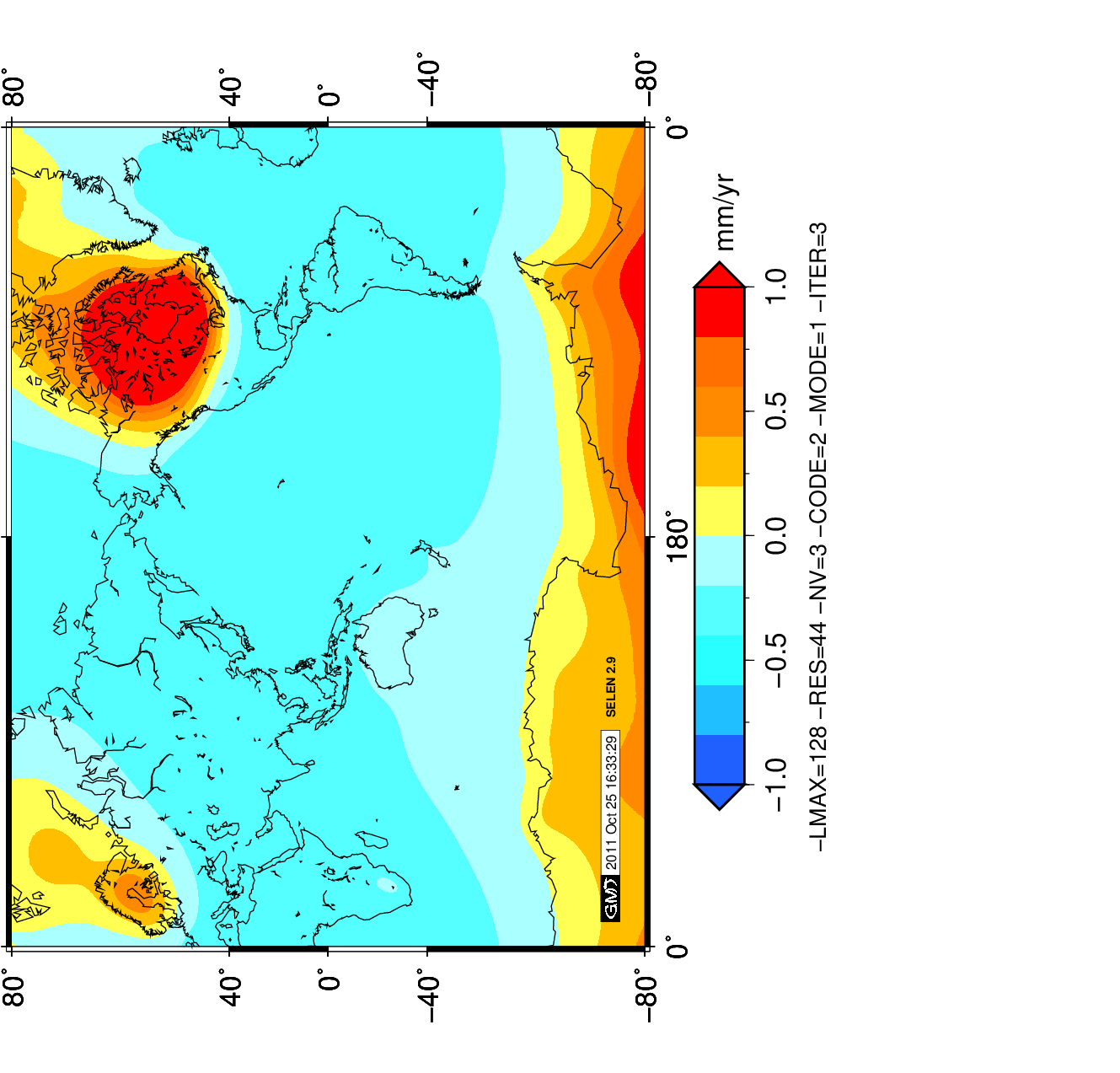}
\caption{Global map showing the sea surface variation induced by GIA, relative 
to the Earth's center of mass ($\dot N$), for run \texttt{TEST}. The range of variation of $\dot N$ 
on this map is $[-0.4/2.3]$ mm/yr.}
\label{fig:ndot}
\end{figure}
\clearpage

\begin{figure}
\begin{center}
\vspace{2cm}
\noindent\includegraphics[width=1.3\textwidth, angle=-90]{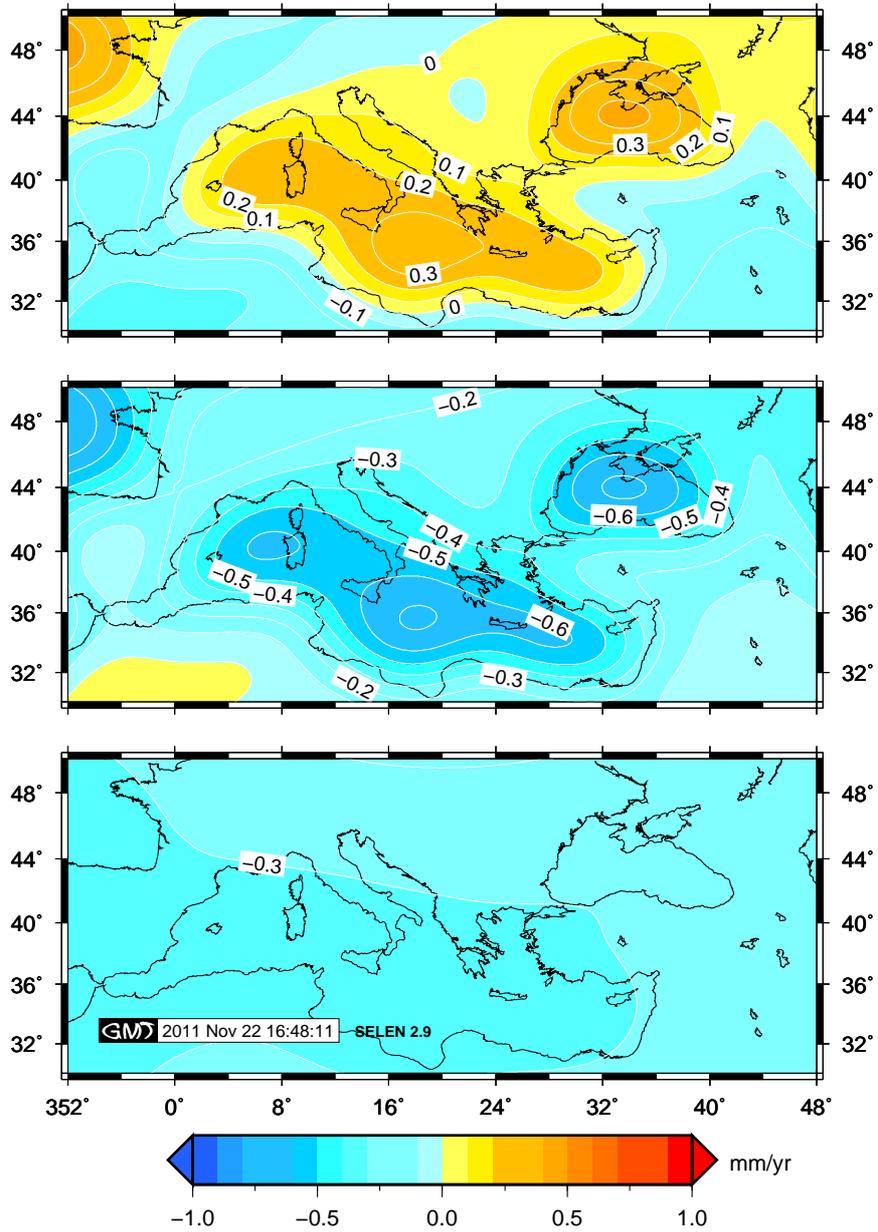}
\vspace{-2cm}
\caption{Regional analysis showing rates of \sealevel change ($\dot S$, top), of vertical uplift 
($\dot U$, middle) and sea surface variation ($\dot N$, bottom) across the Mediterranean region, for run 
\texttt{TEST}. Data and plots for this analysis are found in folder 
\texttt{depot-TEST/rmaps}.}
\label{fig:sun-med}
\end{center}
\end{figure}
\clearpage

\newpage
\begin{figure}
\begin{center}
\vspace{2cm}
\noindent\includegraphics[width=0.8\textwidth, angle=-90]{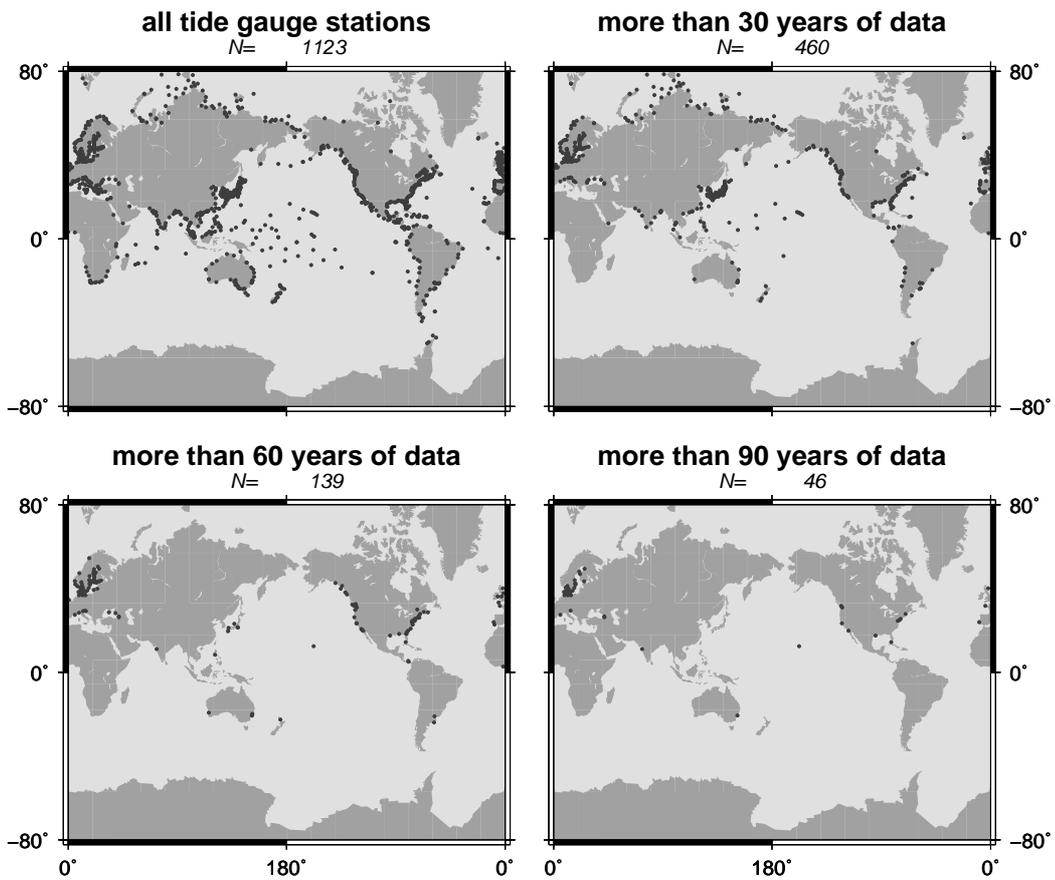}
\caption{Geographical distribution of the TGs considered in run \texttt{TEST}, according to the 
number of RLR data available from each station. Data and plots for this analysis are available 
in \texttt{depot-TEST/ tgauges/ tgauges-sites} after the execution of \selens.}
\label{fig:tg-distribution}
\end{center}
\end{figure}
\clearpage

\begin{table}
\centering
\begin{tabular}{lccccccc}
\hline
Station Name &valid yearly &time range & trend $r_{k}$ & error $\sigma_{k}$ & 
$\displaystyle{r^{gia}_{k}} $ &$\displaystyle{\dot N}$ & $\displaystyle{\dot U}$\\
             &records  & year--year & \multicolumn{2}{c}{mm/yr} & mm/yr &mm/yr &mm/yr\\
\hline
Kungholmsfort &$118$  &  $1887$  --$2005$ & $-0.04$ &  $ 0.13$ & $-1.48$ &   $ +0.03$    &  $+1.52$\\
Olands Norra Udde     &$119$  &  $1887$--$2005$ & $-1.14$ &  $ 0.14$ & $-2.60$ &   $ +0.13$    &  $+2.76$\\
Landsort              &$119$  &  $1887$--$2005$ & $-2.91$ &  $ 0.14$ & $-4.18$ &   $ +0.23$    &  $+4.44$\\
Nedre Sodertaje       &$102$  &  $1869$--$1970$ & $-3.44$ &  $ 0.19$ & $-4.73$ &   $ +0.27$    &  $+5.02$\\
Stockholm             &$117$  &  $1889$--$2005$ & $-3.90$ &  $ 0.15$ & $-4.89$ &   $ +0.28$    &  $+5.20$\\
Ratan                 &$112$  &  $1892$--$2005$ & $-7.81$ &  $ 0.18$ & $-8.95$ &   $ +0.53$    &  $+4.49$\\
Helsinki              &$124$  &  $1879$--$2004$ & $-2.43$ &  $ 0.16$ & $-4.23$ &   $ +0.29$    &  $+4.55$\\
Swinoujscie           &$180$  &  $1811$--$1999$ & $+0.82$ &  $ 0.06$     & $-0.43$ &   $ -0.08$    &  $+0.35$\\
Warnemunde 2          &$149$  &  $1856$--$2005$ & $+1.20$ &  $ 0.07$     & $-0.36$ &   $ -0.09$    &  $+0.27$\\
Wismar 2              &$156$  &  $1849$--$2005$ & $+1.39$ &  $ 0.06$     & $-0.26$ &   $ -0.11$    &  $+0.15$\\
Kobenhavn              &$104$  &  $1889$--$2002$ & $+0.39$ &  $ 0.11$     & $-0.99$ &   $ -0.02$    &  $+0.98$\\
Fredericia             &$105$  &  $1890$--$2002$ & $+1.01$ &  $ 0.08$     & $-0.57$ &   $ -0.06$    &  $+0.52$\\
Aarhus                  &$101$  &  $1889$--$2002$ & $+0.55$ &  $ 0.08$     & $-0.95$ &   $ -0.03$    &  $+0.93$\\
Esbjerg                 &$103$  &  $1890$--$2002$ & $+1.18$ &  $ 0.14$     & $-0.35$ &   $ -0.08$    &  $+0.27$\\
Cuxhaven 2              &$160$  &  $1843$--$2002$ & $+2.44$ &  $ 0.09$     & $-0.01$ &   $ -0.14$    &  $-0.12$\\
Aberdeen II             &$103$  &  $1862$--$1965$ & $+0.58$ &  $ 0.10$     & $-0.43$ &   $ -0.17$    &  $+0.28$\\
Marseille              &$108$  &  $1886$--$2004$ & $+1.28$ &  $ 0.08$     & $+0.08$      &   $ -0.30$    &  $-0.39$\\
Poti                  &$121$  &  $1874$--$2004$ & $+6.55$ &  $ 0.15$     & $+0.03$      &   $ -0.28$    &  $-0.31$\\
Batumi                  &$106$  &  $1882$--$2005$ & $+1.78$ &  $ 0.20$     & $+0.00$    &   $ -0.28$    &  $-0.29$\\
Mumbay/Bombay          &$111$  &  $1878$--$1993$ & $+0.78$ &  $ 0.09$     & $-0.17$     &   $ -0.27$    &  $-0.10$\\
Sydney, F. Denison      &$108$  &  $1886$--$1993$ & $+0.59$ &  $ 0.09$     & $-0.32$    &   $ -0.16$    &  $+0.15$\\
Honolulu              &$101$  &  $1905$--$2005$ & $+1.48$ &  $ 0.12$     & $-0.16$      &   $ -0.35$    &  $-0.18$\\
Seattle              &$106$  &  $1899$--$2005$ & $+2.06$ &  $ 0.10$     & $+0.71$       &   $ -0.02$    &  $-0.73$\\
San Francisco           &$151$  &  $1855$--$2005$ & $+1.43$ &  $ 0.08$     & $+0.68$    &   $ -0.32$    &  $-0.99$\\
Baltimore             &$102$  &  $1903$--$2005$ & $+3.11$ &  $ 0.10$     & $+1.15$      &   $ -0.10$    &  $-1.25$\\
New York           &$131$  &  $1856$--$2005$ & $+2.76$ &  $ 0.06$     & $+1.13$         &   $ -0.05$    &  $-1.19$\\
\hline
\label{table:tg1}
\end{tabular}
\caption{Present--day trends of GIA--induced \sealevel change, sea surface variation, and vertical velocity at 
RLR PSMSL TGs with more than $100$ valid years in the record series. With $r_{k}$ and $\sigma_{k}$ we
denote the best--fit secular rates and their uncertainties. This is an excerpt of the ASCII 
table \texttt{depot-TEST/ tauges/ tgauges-predictions/ ptidegauges}.}
\end{table}
\clearpage

\newpage
\begin{table}
\centering
\begin{tabular}{lccccccc}
\hline
Station Name &valid yearly &time range & trend $r_{k}$ & error $\sigma_{k}$ & 
$\displaystyle{r^{gia}_{k}}$ &$\displaystyle{\dot N}$ & $\displaystyle{\dot U}$\\
             &records  & year--year & \multicolumn{2}{c}{mm/yr} & mm/yr &mm/yr &mm/yr\\
\hline
Algeciras        &$34$   &  $1944$--$2001$ & $+0.46$ &  $ 0.25$     & $-0.21$ &   $ -0.33$    &  $-0.11$\\
Tarifa           &$46$   &  $1944$--$2001$ & $+0.02$ &  $ 0.39$    & $-0.21$ &   $ -0.33$    &  $-0.12$\\
Malaga           &$41$   &  $1994$--$2001$ & $+2.41$ &  $ 0.43$    & $-0.23$               &   $ -0.32$    &  $-0.09$\\
 Alicante II     &$30$   &  $1960$--$1995$ & $-0.85$ &  $ 0.27$ & $-0.07$ &   $ -0.32$    &  $-0.25$\\
Marseille        &$108$  &  $1886$--$2004$ & $+1.28$ &  $ 0.08$     & $+0.08$      &   $ -0.30$    &  $-0.39$\\
Genova           &$78$   &  $1884$--$1992$ & $+1.20$ &  $ 0.07$     & $+0.07$      &   $ -0.29$    &  $-0.36$\\
Venezia (S. Stefano) &$45$      &  $1872$--$1919$ & $+2.55$ &  $ 0.42$     & $+0.04$      &   $ -0.27$    &  $-0.31$\\
Venezia (P. Salute)     &$82$  &  $1909$--$2000$ & $+2.39$ &  $ 0.16$     & $+0.04$      &   $ -0.27$    &  $-0.31$\\
Trieste                 &$96$   &  $1905$--$2006$ & $+1.17$ &  $ 0.12$     & $+0.04$      &   $ -0.27$    &  $-0.31$\\
Rovinj                  &$48$   &  $1956$--$2004$ & $+0.53$ &  $ 0.29$     & $+0.06$      &   $ -0.27$    &  $-0.34$\\
Bakar                   &$62$   &  $1930$--$2004$ & $+0.97$ &  $ 0.23$     & $+0.05$      &   $ -0.27$    &  $-0.32$\\
Split Rt Marjana        &$50$  &  $1953$--$2004$ & $+0.60$ &  $ 0.30$     & $+0.10$      &   $ -0.30$    &  $-0.39$\\
Split Harbour           &$50$  &  $1955$--$2004$ & $+0.33$ &  $ 0.30$     & $+0.10$      &   $ -0.29$    &  $-0.39$\\
Ceuta                   &$51$  &  $1945$--$2005$ & $+0.38$ &  $ 0.23$     & $-0.21$ &   $ -0.33$    &  $-0.11$\\
\hline
\label{table:tg2}
\end{tabular}
\caption{Present--day trends of GIA--induced \sealevel change, sea surface variation, and vertical velocity at 
RLR Mediterranean PSMSL TGs with more than $30$ valid years in the record series.}
\end{table}

\begin{figure}
\begin{center}
\vspace{2cm}
\noindent\includegraphics[width=0.7\textwidth, angle=-90]{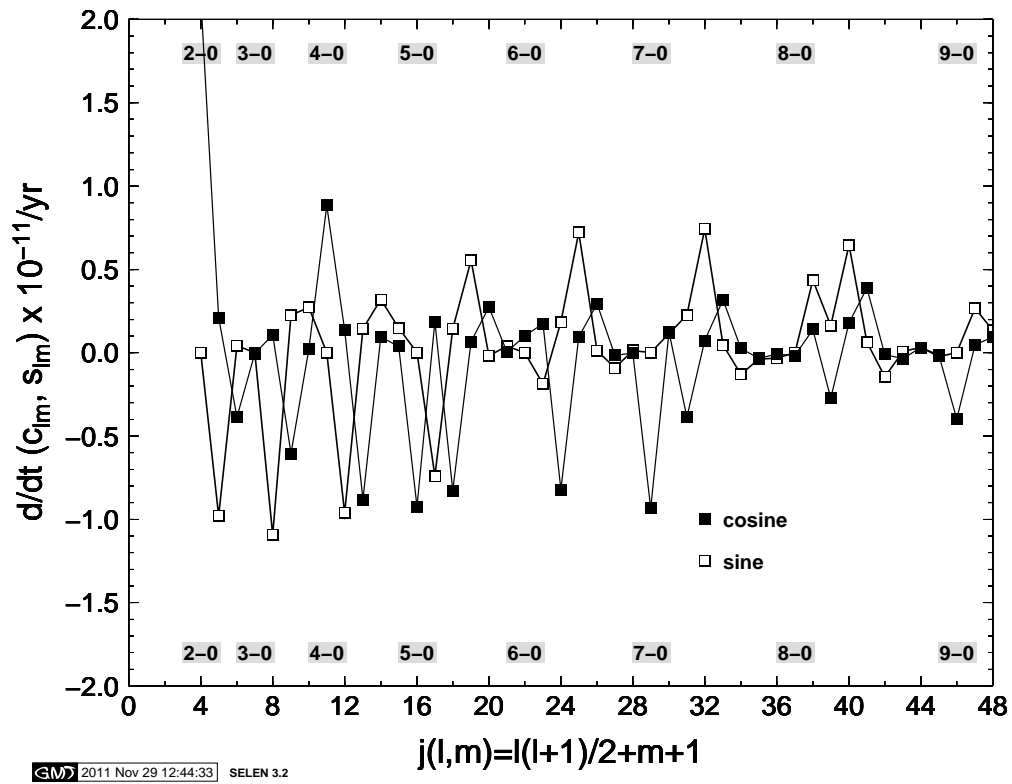}
\caption{Time--derivatives of the Stokes coefficients of the Earth's gravity field associated with GIA, as a function of the generalized harmonic degree $j=\ell(\ell+1)/2+m+1$ for $2\le \ell\le 9$ and $0\le m \le\ell$. Output data for this analysis are found in folder \texttt{depot-TEST/stokes} after the execution of \selens.}
\label{fig:stokes}
\end{center}
\end{figure}
\clearpage


\end{document}